\documentclass[preprint,12pt,authoryear,nopreprintline]{elsarticle}
\usepackage{amsmath,amssymb}
\usepackage{array,booktabs,tabularx}
\usepackage{float,placeins}
\usepackage[hidelinks]{hyperref}
\journal{}
\date{}
\biboptions{round,authoryear}
\makeatletter
\newcommand{\fs@algdouble}{%
  \def\@fs@cfont{\bfseries}%
  \let\@fs@capt\floatc@ruled
  \def\@fs@pre{\hrule height.8pt depth0pt \kern1pt\hrule height.4pt depth0pt \kern2pt}%
  \def\@fs@mid{\kern2pt\hrule height.4pt depth0pt\kern2pt}%
  \def\@fs@post{\kern2pt\hrule height.4pt depth0pt \kern1pt\hrule height.8pt depth0pt}%
  \let\@fs@iftopcapt\iftrue
}
\makeatother
\floatstyle{algdouble}
\newfloat{algorithm}{tbp}{loa}
\floatname{algorithm}{Algorithm}
\newcolumntype{L}{>{\raggedright\arraybackslash}X}
\newcolumntype{C}{>{\centering\arraybackslash}X}
\usepackage[a4paper,margin=25mm]{geometry}

\begin{document}
\begin{frontmatter}
\title{An Automated Framework for Aero-Structural Shape Optimization: Application to Corner-Recessed Steel Arch Sections}
\author[a]{Wenkai Du}
\author[a]{Guangzhong Gao\corref{cor1}}
\ead{ggz@chd.edu.cn}
\author[b]{Jialei He}
\author[a]{Yu Li}
\author[a]{Hua Bai}

\cortext[cor1]{Corresponding author.}
\affiliation[a]{organization={Highway College, Chang'an University},
    city={Xi'an}, postcode={710064}, state={Shaanxi}, country={China}}
\affiliation[b]{organization={School of Future Transportation, Chang'an University},
    city={Xi'an}, postcode={710064}, state={Shaanxi}, country={China}}

\begin{abstract}
Cross-sectional geometry plays a critical role in the wind-resistant design of slender steel structures, governing both aerodynamic excitation and structural resistance. However, aerodynamic modifications aimed at vibration mitigation often conflict with structural stability, while the computational burden of repeated unsteady simulations hinders joint aero-structural optimization. To address these challenges, this paper presents an automated multi-objective aero-structural shape optimization framework. A Python-driven workflow integrates parametric geometry generation, automatic meshing, aeroelastic simulations, and structural finite-element analysis, with sectional and dynamic properties updated for each candidate geometry. Multi-objective Bayesian optimization using Gaussian processes directs sequential batch exploration, while an adaptive wind-speed sampling strategy concentrates evaluations around rapidly varying vibration regimes. The framework is demonstrated on a 200 m-span bare steel arch, simultaneously minimizing the peak vibration amplitude across a prescribed wind-speed range and maximizing the in-plane and out-of-plane compression--bending stability margins. The resulting Pareto front achieves a 25.2\% increase in hypervolume relative to the initial Latin hypercube design set. Crucially, the adaptive wind-speed sampling successfully captures a sharp response peak that uniform sampling misses, while reducing solver core-hours by 69.9\%. The identified trade-offs provide actionable engineering insights into the balanced design of bluff steel structures against aerodynamic instability and structural failure.
\end{abstract}
\end{frontmatter}

\section{Introduction}
\label{sec:1}

Rectangular bluff sections are widely used in bridge members, high-rise buildings, and other slender civil structures. Their aerodynamic behavior is sensitive to the overall sectional proportions and corner geometry, which influence flow separation, shear-layer development, and vortex shedding \citep{Kawai1998}. For flexible structures---including arch ribs and hangers---vortex-induced vibration (VIV) and galloping can generate repeated or large-amplitude oscillations, raising serviceability and fatigue concerns \citep{Li2024,Gao2016,Gao2020}. Wind-tunnel investigations of long-span arches further demonstrate the need to assess aeroelastic stability during design \citep{Mannini2016,Chen2025}. Sectional modifications offer opportunities to improve wind resistance: corner cuts and recessions can modify aerodynamic damping, while rounded corners alter drag and shedding characteristics \citep{Hnin2024,Powar2026}. These findings motivate systematic sectional optimization guided by the relevant vibration-response measures.

Aerodynamic shape optimization has enabled systematic tailoring of bluff-body sections. \citet{Bernardini2015} optimized building cross-sections using lift and drag coefficients as competing objectives. \citet{Elshaer2017} developed an aerodynamic corner-modification procedure for tall buildings, accounting for turbulence and wind directionality and verifying selected geometries through wind-response calculations. More recently, \citet{Das2023} optimized the corner-recessed geometry of a square building to reduce mean wind pressures. These studies confirm the relevance of sectional proportions and corner geometry as design variables. However, force coefficients and surface pressures characterize aerodynamic performance differently from the vibration amplitude of an elastically supported section. Improvements in these static aerodynamic quantities do not by themselves establish how a modified section will respond dynamically when its mass, stiffness, and natural frequency change simultaneously.

Structural integrity imposes complementary constraints, as the identical cross-sectional geometry simultaneously dictates material expenditure, directional stiffness, and load-bearing capacity. For steel arches, geometric optimization has been used to reduce material consumption while satisfying strength constraints \citep{Korus2021}. When wind-induced response governs, these structural effects must be considered together with aerodynamic changes. Bridge-deck optimization provides examples of this coupling: \citet{Zheng2023} incorporated geometry-dependent dynamic properties into flutter optimization, while \citet{CidMontoya2021,CidMontoya2022} addressed structural and aeroelastic requirements jointly through combined shape and size optimization. A subsequent sequential formulation linked structural sizing with aeroelastic shape optimization to improve computational efficiency \citep{CidMontoya2024}. For wind-sensitive structures, integrated aero-structural optimization is therefore essential to identify sectional geometries that balance vibration control with structural stability.

\begin{figure}[tbp]
\centering
\includegraphics[width=0.98\linewidth,height=0.73\textheight,keepaspectratio]{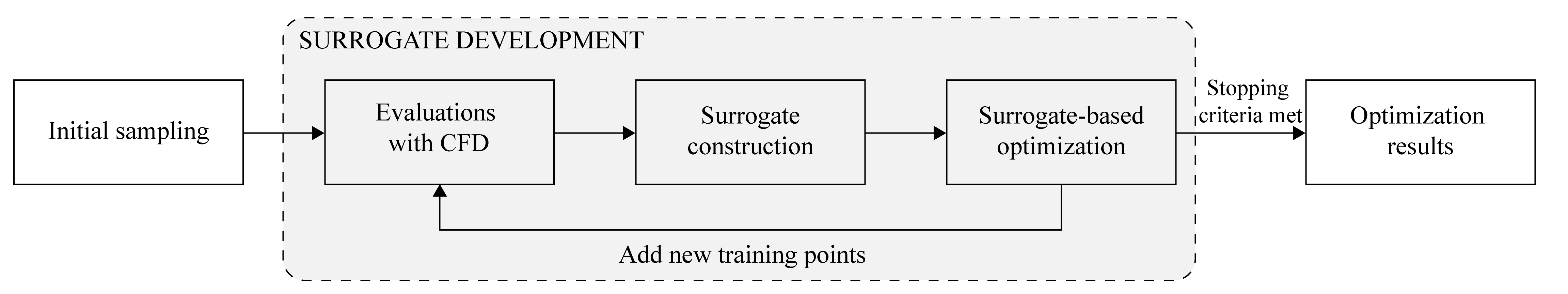}
\caption{General workflow of surrogate-assisted shape optimization.}\label{fig:1}
\end{figure}

Such optimization requires efficient performance evaluation and design-space exploration. Empirical modifications and wind-tunnel studies reveal useful trends, but the limited number of configuration comparisons cannot fully represent interactions among geometric variables, and exhaustive parametric scans become prohibitively costly (\citealp{Jafari2021}; Li et al., 2023). Parametric modelling and computational fluid dynamics (CFD) allow candidate geometries to be generated and assessed consistently, yet repeated unsteady flow and fluid--structure interaction (FSI) simulations remain expensive. Surrogate models alleviate this cost by approximating the mapping from design variables to responses. Relevant approaches include adaptive Kriging models for building-section optimization \citep{Bernardini2015}, neural-network surrogates for corner modification \citep{Elshaer2017}, and polynomial emulators for bridge-deck design \citep{Tinmitond2022}. The accuracy of these surrogates depends on the available training data and on where additional evaluations are placed. In this context, Bayesian optimization uses probabilistic predictions and their uncertainty to guide sequential sampling. Efficient global optimization (EGO) and its multi-objective extension ParEGO provide established methodological foundations \citep{Jones1998,Knowles2006}, with recent applications to constrained bridge design \citep{Rostum2025}. Fig.~\ref{fig:1} summarizes the resulting cycle of surrogate construction, design selection, direct evaluation, and model update.

For vibration-based shape optimization, selecting the next geometry addresses only part of the evaluation cost. Each geometry should also be assessed at the wind speeds needed to characterize its response adequately. When the aerodynamic objective is the maximum vibration amplitude over a prescribed range, changes in sectional geometry and natural frequency can shift the location and magnitude of the response peak, while VIV--galloping interactions may further complicate the response curve \citep{Mannini2014}. Dense wind-speed scans increase the per-design computational cost, whereas sparse scans risk missing narrow peaks and underestimating the vibration objective. Effective optimization therefore involves two coupled sampling decisions: which geometry to evaluate next, and which wind speeds to evaluate for that geometry. Addressing both is necessary to combine efficient design-space exploration with adequate resolution of the sampled vibration envelope.

This study develops an automated multi-objective aero-structural optimization framework. The framework connects CFD--FSI simulations and three-dimensional structural finite-element analysis through shared geometric variables and physical data exchange. It is demonstrated on a single bare steel arch in the construction stage, with normalized vibration amplitude and compression--bending stability margin serving as the competing objectives. The main contributions are as follows.

\begin{enumerate}

\item An adaptive wind-speed sampling strategy that combines frequency-informed initialization with response-guided refinement to evaluate the vibration envelope within a prescribed sampling budget. A comparison with uniform sampling demonstrates its ability to capture a narrow response peak that uniform sampling misses, while requiring fewer evaluations and lower solver cost.

\item An automated aero-structural optimization framework that integrates this wind-speed sampling strategy with multi-objective Bayesian optimization for geometric design selection. For each candidate, the sectional and modal properties derived from the structural model inform the FSI evaluation, while the aerodynamic coefficients obtained from CFD enter the structural wind-load FEM model and the stability assessment.

\item A systematic investigation of corner-recessed rectangular sections applied to a steel arch, revealing how the aspect ratio and recession dimensions jointly affect the aerodynamic response, structural stability, and their trade-offs. The observed Pareto front and associated computational costs are analyzed to support engineering selection among vibration suppression, stability, and material economy.

\end{enumerate}

The remainder of this paper is organized as follows. Section~\ref{sec:2} presents the methodology and implementation. Section~\ref{sec:3} defines the application case, optimization problem, and numerical settings. Section~\ref{sec:4} discusses the optimization results and examines the effects of sectional geometry on aerodynamic and structural performance. Finally, Section~\ref{sec:5} summarizes the conclusions.

\section{Proposed aero-structural shape optimization framework}
\label{sec:2}

This section presents an automated framework for multi-objective aero-structural shape optimization. Section~\ref{sec:2.1} outlines the overall architecture, and Section~\ref{sec:2.2} introduces the adaptive wind-speed sampling strategy. Sections~\ref{sec:2.3}-2.5 describe mesh generation and the aerodynamic and structural evaluation methods. Section~\ref{sec:2.6} presents the multi-objective Bayesian optimization method for sequential candidate selection, followed by the software implementation in Section~\ref{sec:2.7}.

\subsection{Aero-structural shape optimization framework}
\label{sec:2.1}

\begin{figure}[tbp]
\centering
\includegraphics[width=0.98\linewidth,height=0.73\textheight,keepaspectratio]{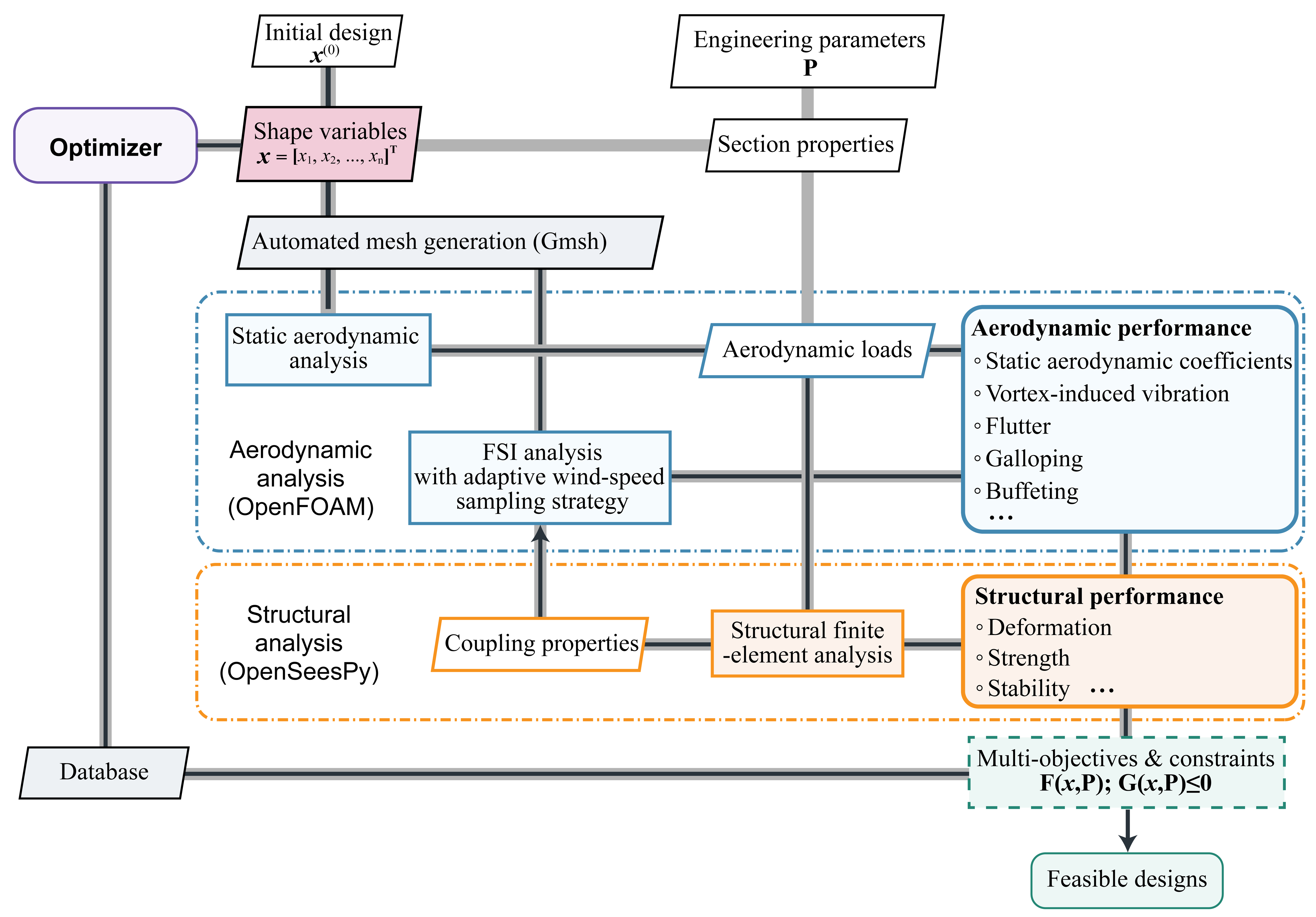}
\caption{Proposed multi-objective aerodynamic-structural shape optimization framework}\label{fig:2}
\end{figure}

The proposed framework integrates aerodynamic and structural performance evaluations within a common shape-optimization loop, as illustrated in Fig.~\ref{fig:2}. A shared set of shape variables defines the geometry used in both aerodynamic and structural analyses. Each candidate is evaluated through the corresponding numerical models, and the computed responses are converted into objective and constraint values. These values are recorded in a database and returned to the optimizer to guide subsequent candidate selection. The framework thus links geometric modification, coupled performance evaluation, and sequential optimization in an automated workflow.

The shape is represented by the design vector \(x=[x_{1},x_{2},\ldots ,x_{n}]^{\text{T}}\), where \(n\) is the number of design variables. A separate parameter set, \(\boldsymbol{p}\), specifies quantities that remain fixed during optimization, such as material properties, overall structural dimensions, boundary conditions, and environmental conditions. The general optimization problem is expressed as

\begin{equation}
\begin{alignedat}{2}
&\operatorname*{minimize}_{x} &&\quad \boldsymbol{F}(x;\boldsymbol{p})
=\left[f_{1}(x;\boldsymbol{p}),\ldots,f_{m}(x;\boldsymbol{p})\right]^{\text{T}}\\
&\text{subject to} &&\quad \boldsymbol{G}(x;\boldsymbol{p})\leq 0\\
&&&\quad x_{\text{L}}\leq x\leq x_{\text{U}}
\end{alignedat}
\label{eq:1}
\end{equation}
where \(\boldsymbol{F}\) contains the \(m\) selected objectives, \(\boldsymbol{G}\) refers to the prescribed constraints, \(x_{\text{L}}\) and \(x_{\text{U}}\) define the lower and upper bounds of the design variables. Maximization objectives are transformed to the minimization convention where required. Aerodynamic and structural response quantities can be assigned as objectives or constraints according to the engineering requirements, allowing the formulation to accommodate different design priorities.

For each design point, the parameterized geometry is regenerated and its section properties are updated. These properties and the engineering parameters define the structural model. The aerodynamic computational mesh is generated automatically to maintain consistency with the modified shape. In the present implementation, Gmsh generates the computational mesh, OpenFOAM performs the static aerodynamic and fluid-structure interaction analysis, and OpenSeesPy conducts the structural finite-element analysis. These software components communicate through the execution and data-processing layer described in Section~\ref{sec:2.7}.

The aerodynamic and structural evaluations are connected through the exchange of physical parameters. Fixed-body simulations provide aerodynamic coefficients for defining structural wind loads, while structural modal properties determine the dynamic parameters of the FSI model. The FSI simulations then evaluate wind-induced vibration at selected wind speeds, with adaptive sampling refining the vibration-response envelope. In parallel, the structural analysis evaluates deformation and internal forces under the prescribed loads, from which strength and stability measures are derived. Consequently, each shape modification is assessed through its effects on both aerodynamic loading and structural properties, including mass, stiffness, and dynamic characteristics.

The resulting objective and constraint values are returned to the optimization module to select the next batch of candidates. Although the modular architecture accommodates different optimization algorithms, the present implementation adopts multi-objective Bayesian optimization with Gaussian-process surrogates. Newly evaluated designs enrich the database and update the surrogate models, forming a sequential cycle of prediction, candidate selection, and direct numerical evaluation. Once the stopping criteria are satisfied, the observed non-dominated designs are extracted to characterize aero-structural trade-offs and support the selection of representative engineering solutions.

\subsection{Adaptive wind-speed sampling strategy}
\label{sec:2.2}

Evaluating vortex-induced vibration (VIV) over a prescribed wind-speed range requires multiple transient fluid-structure interaction simulations for each geometry. Uniform sampling may allocate many evaluations to low-response regions while leaving a narrow response peak insufficiently sampled. An adaptive wind-speed sampling strategy is therefore introduced to concentrate additional evaluations near the controlling sampled response and intervals with substantial amplitude variation. It combines frequency-informed initialization, response-driven refinement, and budget-controlled termination, as summarized in Algorithm~\ref{alg:1}.

\begin{algorithm}[tbp]
\caption{Adaptive wind-speed sampling strategy}\label{alg:1}
\small
\smallskip
\noindent\textbf{Input:} natural frequency \(f_n\), characteristic dimension \(D\), Strouhal number \(St_D\), reference wind speed \(U_{\text{ref}}\); budgets \(K_{\max}\) refinement rounds, \(m_{\max}\) additions per round, and \(N_{\max}\) total wind speeds; thresholds \(\theta_A\), \(\tau_d\), and \(\tau_A\).\par
\noindent\textbf{Output:} sampled response envelope \(A/D\), sampled wind-speed set \(S\), and quality flags \(Q\).\par
\smallskip
\hrule
\begin{list}{}{%
  \setlength{\leftmargin}{2.7em}%
  \setlength{\labelwidth}{2.2em}%
  \setlength{\labelsep}{0.5em}%
  \setlength{\itemsep}{0pt}%
  \setlength{\parsep}{0pt}%
  \setlength{\topsep}{0.3em}%
}
\item[] \textit{\textbf{Initialization}}
\item[1:] \(U^{*}\leftarrow f_nD/St_D;\ Q\leftarrow\varnothing\)
\item[2:] \textbf{if} \(U^{*}\leq U_{\text{ref}}\) \textbf{then}
\item[3:] \(\quad S\leftarrow\{0.70,0.90,1.10,1.30\}U^{*}\cup\{(U^{*}+U_{\text{ref}})/2,U_{\text{ref}}\}\)
\item[4:] \textbf{else} \(S\leftarrow\{iU_{\text{ref}}/6:i=1,\ldots,6\}\); flag the above-limit anchor
\item[5:] \textbf{end if}
\item[6:] Clip \(S\) to \((0,U_{\text{ref}}]\), sort and deduplicate; fill any shortfall by bisecting the largest gaps
\item[7:] \(R\leftarrow\) successful evaluations of \(a(U)=A(U)/D\) at \(S\); \textbf{if} \(R=\varnothing\) \textbf{then} report failure; \textbf{return}
\item[8:] \(a_{\max}^{(0)}\leftarrow\max_{R}a\)
\item[] \textit{\textbf{Adaptive refinement}}
\item[9:] \textbf{for} \(k=1,\ldots,K_{\max}\) \textbf{do}
\item[10:] \textbf{if} \(|S|\geq N_{\max}\) \textbf{then} set budget flag; \textbf{break}
\item[11:] \(C_b\leftarrow\) admissible boundary-extension candidates; \(C_I\leftarrow\varnothing\)
\item[12:] \textbf{for each} adjacent pair \((U_i,a_i)\) and \((U_j,a_j)\) in sorted \(R\) \textbf{do}
\item[13:] \(\quad d_{ij}\leftarrow |a_i-a_j|/\max(a_i,a_j)\)
\item[14:] \(\quad\) \textbf{if} \(\max(a_i,a_j)\geq\theta_Aa_{\max}^{(k-1)}\) and \(d_{ij}>\tau_d\) \textbf{then}
\item[15:] \(\qquad s_{ij}\leftarrow[\max(a_i,a_j)/a_{\max}^{(k-1)}]d_{ij}\); add \(((U_i+U_j)/2,s_{ij})\) to \(C_I\)
\item[16:] \hspace*{1em}\textbf{end if; end for}
\item[17:] \(C\leftarrow\) up to \(\min(m_{\max},N_{\max}-|S|)\) new speeds: \(C_b\) first, then \(C_I\) in decreasing \(s_{ij}\)
\item[18:] \textbf{if} \(C=\varnothing\) \textbf{then} record the stopping flag; \textbf{break}
\item[19:] \(S\leftarrow S\cup C;\ R\leftarrow R\cup\) successful evaluations at \(C\)
\item[20:] \(a_{\max}^{(k)}\leftarrow\max_R a;\ \delta_A\leftarrow|a_{\max}^{(k)}-a_{\max}^{(k-1)}|/a_{\max}^{(k)}\)
\item[21:] Update peak-bracketing, upper-limit, and remaining-candidate flags
\item[22:] \textbf{if} \(\delta_A\leq\tau_A\) and (peak bracketed or upper-limit-censored) and no candidate remains \textbf{then}
\item[23:] \hspace*{1em}set convergence/censor flag; \textbf{break}
\item[24:] \textbf{else if} \(k=K_{\max}\) or \(|S|\geq N_{\max}\) \textbf{then} set budget flag; \textbf{break}
\item[25:] \textbf{end if; end for}
\item[26:] \(A/D\leftarrow\max_R a\); \textbf{return} \(A/D,S,Q\)
\end{list}
\end{algorithm}

First, the initial samples are guided by the frequency-matching wind speed

\begin{equation}
St=\frac{f_{s}D}{U_{\text{ref}}}
\label{eq:2}
\end{equation}

\begin{equation}
U^{*}=\frac{f_{n}D}{St}
\label{eq:3}
\end{equation}
where \(f_{s}\) is the shedding frequency obtained from a fixed-body simulation at the reference wind speed \(U_{\text{ref}}\), \(f_{n}\) is the structural frequency, and \(D\) is the characteristic dimension. All quantities are expressed at the same physical scale. When \(U^{*}\leq U_{ref}\), six initial speeds are specified as

\begin{equation}
S_{\text{raw}}=\left\{ 0.70U^{*},\,0.90U^{*},\,1.10U^{*},\,1.30U^{*},\,\frac{U^{*}+U_{ref}}{2},\,U_{ref}\right\}
\label{eq:4}
\end{equation}

This arrangement concentrates four samples around the frequency-matching estimate and places two additional samples towards the reference wind speed. The speeds are clipped to the prescribed bounds and duplicate values are removed; any shortfall is filled by successively bisecting the largest remaining gaps until six distinct speeds are obtained. If \(U^{*}>U_{ref}\), six uniformly spaced speeds are used instead. The frequency-matching estimate guides the initial allocation without prescribing the boundaries of the lock-in region.

Next, the normalized response is expressed as \(a\left( U\right) =A\left( U\right) /D\), where \textit{A} is the vibration amplitude. At the beginning of refinement round \textit{k}, the successful evaluations form the dataset \(R_{k-1}\) with the sampled maximum \(a_{max}^{(k-1)}\). For adjacent sampled speeds \(U_{i}<U_{j}\) with \(max\left( a_{i},a_{j}\right) >0\), the relative amplitude difference and refinement score are defined as

\begin{equation}
d_{ij}=\frac{|a_{i}-a_{j}|}{max(a_{i},a_{j})}
\label{eq:5}
\end{equation}

\begin{equation}
s_{ij}=\frac{max(a_{i},a_{j})}{a_{max}^{(k-1)}}d_{ij}
\label{eq:6}
\end{equation}

Here, \(d_{ij}\) measures the relative response variation across an interval, while \(s_{ij}\) weights this variation by its amplitude relative to the current sampled maximum. An interval generates a midpoint candidate\(U_{\text{new}}=(U_{i}+U_{j})/2\), when

\begin{equation}
max(a_{i},a_{j})\geq \theta _{A}a_{max}^{(k-1)}\&d_{ij}>\tau _{\text{d}},
\label{eq:7}
\end{equation}
where \(\theta _{A}\) defines the amplitude threshold for refinement and \(\tau _{\text{d}}\) specifies the required relative variation. Eligible midpoint candidates are ranked by decreasing \(s_{ij}\), directing evaluations towards intervals with both appreciable response and substantial variation. When the current maximum lies at an endpoint of the sampled wind-speed set, admissible outward extensions take priority over midpoint refinement. All extensions remain within the prescribed bounds.

Finally, each refinement round evaluates at most \(m_{\text{max}}\) new wind speeds, subject to a maximum of \(K_{\text{max}}\) refinement rounds and \(N_{\text{max}}\) distinct requested speeds. Convergence is assessed from the relative change in the sampled maximum between successive rounds using the prescribed tolerance \(\tau _{\text{A}}\). The final aerodynamic response measure is the maximum over all successful evaluations,

\begin{equation}
\frac{A}{D}=\operatorname*{max}\limits_{(U_{i},a_{i})\in R_{\mathrm{final}}}a_{i}
\label{eq:8}
\end{equation}

This value is returned together with the sampled response curve, evaluated wind speeds and termination status, providing both the vibration objective and a record of its evaluation.

\begin{figure}[tbp]
\centering
\includegraphics[width=0.98\linewidth,height=0.73\textheight,keepaspectratio]{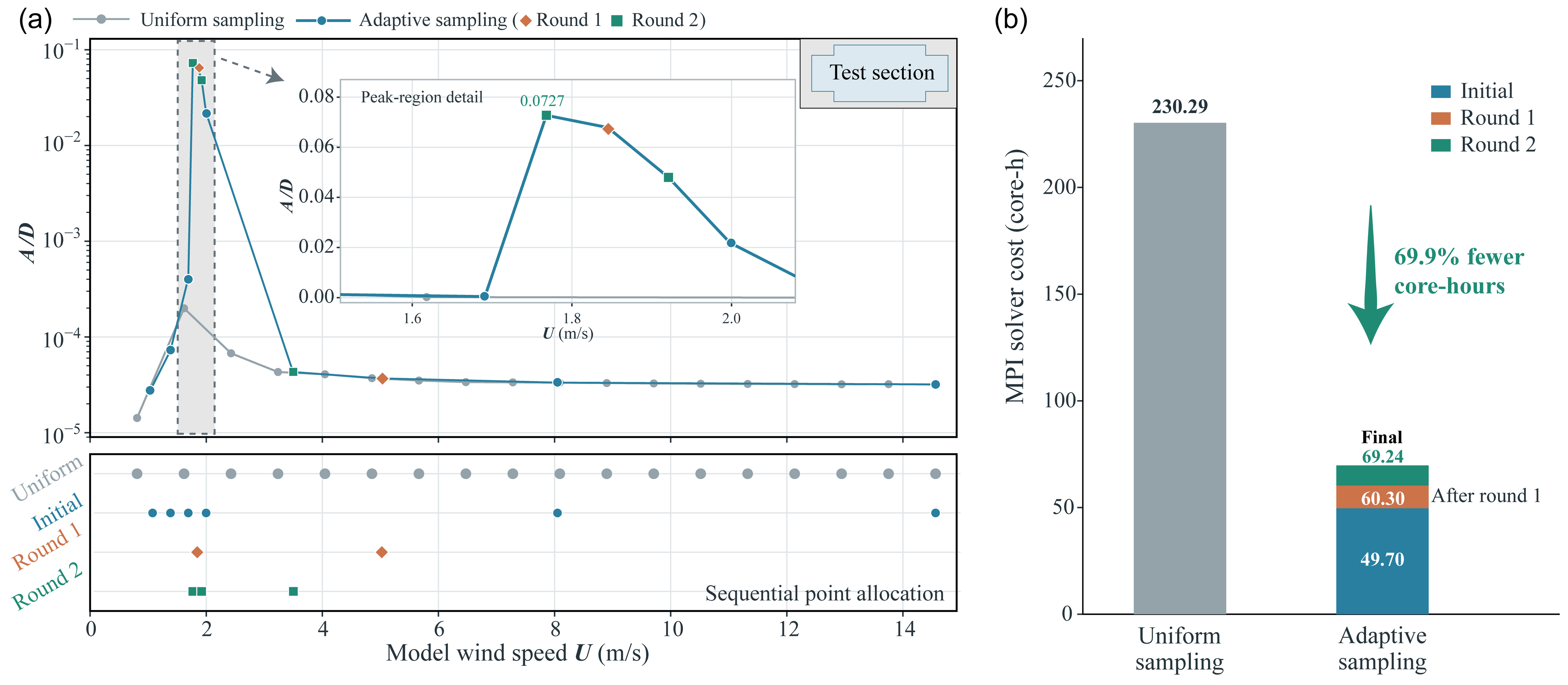}
\caption{A representative example of the adaptive wind-speed sampling strategy: (a) vibration-response curves and sequential wind-speed allocation; (b) comparison of MPI solver core-hours between adaptive and uniform sampling.}\label{fig:3}
\end{figure}

Fig.~\ref{fig:3} demonstrates the effectiveness of the adaptive wind-speed sampling strategy for a representative section. Starting from six initial wind speeds, two refinement rounds add five additional samples near the narrow response peak. As shown in Fig.~\ref{fig:3}(a), the adaptive procedure identifies a sampled maximum of \(A/D\approx 0.0727\) at \(U\approx 1.77\) m/s, which is not captured by the 18-point uniform scan. The allocation diagram further illustrates how additional evaluations are concentrated in the peak region. Meanwhile, the recorded MPI solver cost decreases from 230.29 to 69.24 core-hours, corresponding to a reduction of 69.9\%, as shown in Fig.~\ref{fig:3}(b). In this example, adaptive sampling therefore resolves a higher response peak using 11 rather than 18 wind-speed evaluations, while substantially reducing computational cost.

\subsection{Automated mesh generation and wall function method}
\label{sec:2.3}

An automated mesh generation procedure is established through the Gmsh Python API \citep{Geuzaine2009} to convert geometries into CFD-ready meshes. Gmsh is an open-source mesh generator with geometric modeling and programmable meshing capabilities. The procedure integrates parametric geometry construction, mesh-size control, boundary identification, format conversion, and quality acceptance into a repeatable workflow comprising the following six steps.

\textbf{Step} \textbf{1}\textbf{:} Initialization and geometry construction. Gmsh is initialized, and the sectional contour is constructed from the design variables.

\textbf{Step} \textbf{2}\textbf{:} Computational domain and meshing strategy. The surrounding fluid domain is defined, and an unstructured mesh is generated with local refinement near the body and progressively coarser resolution toward the far field.

\textbf{Step} \textbf{3}\textbf{:} Physical-boundary identification. Geometric and topological relationships are used to identify the fluid region and its boundaries. Consistent physical names are assigned to the inlet, outlet, body wall, and remaining domain boundaries.

\textbf{Step} \textbf{4}\textbf{:} Mesh generation and export. The mesh is generated and exported in the \textit{.}\textit{msh} format.

\textbf{Step} \textbf{5}\textbf{:} Conversion to OpenFOAM. The mesh is converted into the OpenFOAM format using \textit{gmshToFoam}, after which the required boundary-patch types are assigned.

\textbf{St}\textbf{e}\textbf{p} \textbf{6}\textbf{:} Mesh-quality acceptance. Mesh validity, cell volumes, non-orthogonality, and skewness are checked against prescribed criteria. Meshes that fail these checks return to Step 2 for adjustment and regeneration. Only accepted meshes proceed to the flow solution.

For near-wall treatment, the Spalding continuous wall function is adopted through \textit{nutUSpaldingWallFunction} in OpenFOAM. Without wall functions, resolving the viscous sublayer requires a fine near-wall mesh with \(y^{+}<1\) \citep{Bruno2008}, resulting in a higher computational cost. However, wall function method evaluates the turbulent viscosity using a continuous wall-law relation that connects the viscous sublayer, buffer layer, and logarithmic region. Spalding's law is expressed as follows:

\begin{equation}
y^{+}=u^{+}+\frac{1}{E}\left[ exp(\kappa u^{+})-1-\kappa u^{+}-\frac{(\kappa u^{+})^{2}}{2}-\frac{(\kappa u^{+})^{3}}{6}\right]
\label{eq:9}
\end{equation}
where \(u^{+}=U_{t}/u_{\tau}\), \(y^{+}=yu_{\tau}/\nu\), and \(u_{\tau}=\sqrt{|\tau _{w}|/\rho}\). Here, \(U_{t}\) is the magnitude of the wall-parallel velocity relative to the wall at the first cell center, \textit{y} is its wall-normal distance, \(\nu\) is the kinematic viscosity, \(\tau _{w}\) is the wall shear stress, \(\rho\) is the fluid density. The parameters \(\kappa\) and \textit{E} are the von Kármán constant and an empirical wall-law constant, respectively. By providing a continuous transition between near-wall regions, this formulation offers greater flexibility in wall-adjacent mesh placement. Where adequate response accuracy is maintained, it permits coarser near-wall meshes than a fully wall-resolved treatment targeting \(y^{+}\approx 1\), thereby reducing the cell count and the cost of repeated CFD evaluations.

\begin{figure}[tbp]
\centering
\includegraphics[width=0.98\linewidth,height=0.73\textheight,keepaspectratio]{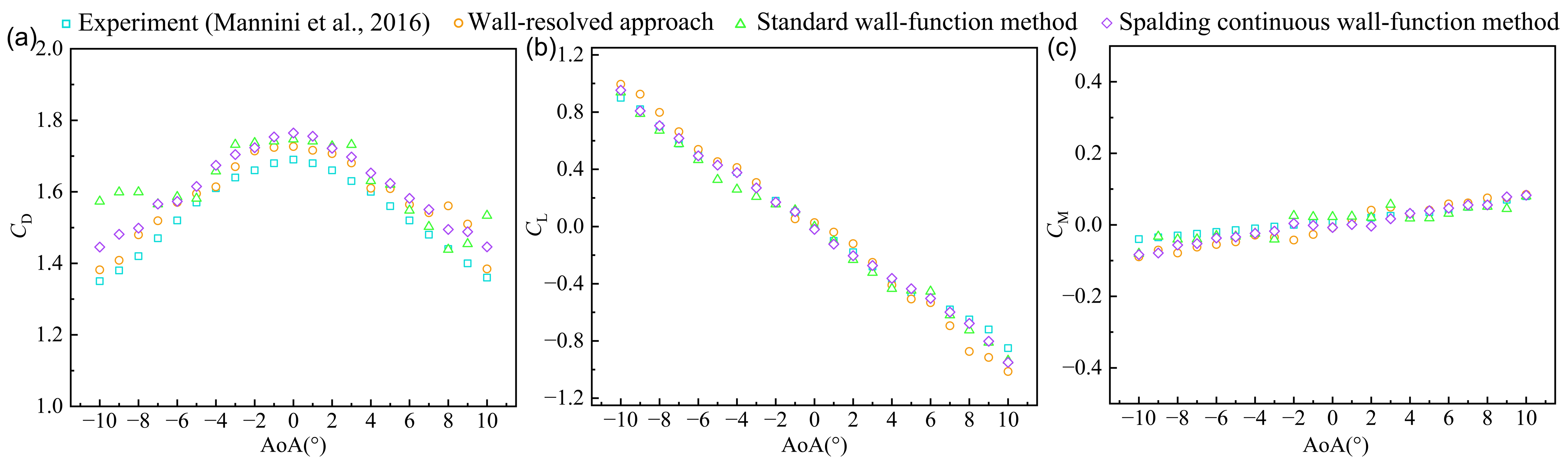}
\caption{Comparison of aerodynamic coefficients obtained using different near-wall treatments methods: (a) drag coefficient \(C_{\text{D}}\); (b) lift coefficient \(C_{\text{L}}\); (c) pitching-moment coefficient \(C_{\text{M}}\).}\label{fig:4}
\end{figure}

Fig.~\ref{fig:4} compares the drag, lift, and pitching-moment coefficients of 3:2 rectangular section obtained using the Spalding treatment with experimental data \citep{Mannini2016}, wall-resolved approach and standard wall-function method. The Spalding-based results show closer agreement with the experimental data at selected incidences than the classical treatment. This comparison supports the adopted near-wall treatment as a viable and effective approach for the benchmark case. Mesh independence should nevertheless be established for the application before conducting the production CFD evaluations, as examined in Section~\ref{sec:3.3.1}.

\subsection{Aerodynamic analysis with OpenFOAM}
\label{sec:2.4}

Aerodynamic analyses are performed using the finite-volume framework of OpenFOAM. The flow is modelled using the incompressible unsteady Reynolds-averaged Navier-Stokes equations (URANS), with turbulence closure provided by the \textit{k-\ensuremath{\omega}} shear-stress transport (\textit{k-\ensuremath{\omega}} SST) model. Two complementary solver configurations are employed: \textit{pisoFoam} for fixed-body simulations and \textit{pimpleFoam} for moving-mesh fluid-structure interaction simulations \citep{Robertson2015,Zhang2023}. Both procedures share the same governing equations, while differing in their pressure-velocity coupling iterations and treatment of mesh motion.

A unified description of the fixed and moving meshes is obtained using the arbitrary Lagrangian-Eulerian (ALE) formulation \citep{Mohammadian2023}. The continuity and momentum equations are expressed as

\begin{equation}
\nabla \cdot \boldsymbol{u}=0
\label{eq:10}
\end{equation}

\begin{equation}
\left. \frac{\partial \boldsymbol{u}}{\partial t}\right\rvert _{\boldsymbol{\chi}}+\left[ (\boldsymbol{u}-\boldsymbol{u}_{g})\cdot \nabla\right] \boldsymbol{u}=-\nabla \varpi +\nabla \cdot \left[ \nu _{\text{eff}}\left( \nabla \boldsymbol{u}+\nabla \boldsymbol{u}^{\text{T}}\right)\right]
\label{eq:11}
\end{equation}
where \(\boldsymbol{u}\) is the Reynolds-averaged fluid velocity, \(\boldsymbol{u}_{g}\) is the mesh velocity, and \(\boldsymbol{\chi}\) denotes the reference mesh coordinates held fixed in the time derivative. The effective kinematic viscosity is \(\nu _{\text{eff}}=\nu +\nu _{t}\), where \(\nu\) and \(\nu _{t}\) are the molecular and turbulent kinematic viscosities, respectively. The pressure variable \(\varpi\) incorporates the isotropic contribution of the modeled Reynolds stresses. For a stationary mesh, \(\boldsymbol{u}_{g}=0\), and the equations reduce to the conventional Eulerian form. For a moving mesh, convection transport is evaluated using the fluid velocity relative to the mesh motion.

After spatial and temporal discretization, the volume-normalized momentum equation for cell \textit{P} is written as

\begin{equation}
a_{P}\boldsymbol{u}_{P}=\boldsymbol{H}_{P}-(\nabla \varpi )_{P}
\label{eq:12}
\end{equation}

\begin{equation}
r_{A,P}=\frac{1}{a_{P}}
\label{eq:13}
\end{equation}

\begin{equation}
\boldsymbol{h}_{P}=r_{A,P}\boldsymbol{H}_{P}
\label{eq:14}
\end{equation}
where \(a_{P}\) is the diagonal momentum coefficient and \(\boldsymbol{H}_{P}\) collects the remaining non-pressure contributions, including neighboring-cell terms and known temporal and source terms. Combining this relation with discrete continuity gives the pressure equation and velocity correction:

\begin{equation}
\sum_{f\in \partial P} (r_{A})_{f}|\boldsymbol{S}_{f}|\left( \frac{\partial \varpi}{\partial n}\right) _{f}=\sum_{f\in \partial P} \Phi _{H,f}
\label{eq:15}
\end{equation}

\begin{equation}
\boldsymbol{u}_{P}=\boldsymbol{h}_{P}-r_{A,P}(\nabla \varpi )_{P}
\label{eq:16}
\end{equation}
where \(\partial P\) denotes the faces of cell \textit{P}, \(\boldsymbol{S}_{f}\) is the outward face-area vector, and \(\Phi _{H,f}\) is the provisional absolute volumetric flux constructed from \textbf{h}, including the required interpolation and temporal corrections. The face-normal pressure gradient is evaluated using the selected discretization, with non-orthogonal corrections where required. The corrected absolute and mesh-relative fluxes are then obtained as

\begin{equation}
\Phi _{f}^{\text{abs}}=\Phi _{H,f}-(r_{A})_{f}|\boldsymbol{S}_{f}|\left( \frac{\partial \varpi}{\partial n}\right) _{f}
\label{eq:17}
\end{equation}

\begin{equation}
\Phi _{f}^{\text{rel}}=\Phi _{f}^{\text{abs}}-\Phi _{g,f}
\label{eq:18}
\end{equation}
where \(\Phi _{g,f}\) is the mesh-motion flux. The pressure correction enforces continuity using the absolute flux, whereas the relative flux describes convective transport across moving cell faces. These operations provide a common representation of the pressure-velocity coupling used by both solvers.

In \textit{pisoFoam}, the Pressure-Implicit with Splitting of Operators (PISO) algorithm performs a momentum prediction followed by pressure-velocity corrections within each physical time step \citep{Fernandes2026}. This configuration is used to obtain mean aerodynamic coefficients and vortex-shedding frequencies for stationary sections. Although the body and mesh remain fixed, the flow solution is transient, allowing unsteady vortex shedding to be resolved.

In \textit{pimpleFoam}, the PIMPLE algorithm combines PISO-type pressure-velocity corrections with SIMPLE-type outer iterations. The outer loop revisits the momentum and pressure equations within each time step, allowing the nonlinear coupling to be updated repeatedly \citep{Mathupriya2018}. A single outer iteration produces a PISO-like procedure, whereas additional outer iterations permit further within-step correction. In the present framework, this configuration is combined with body-motion integration and mesh deformation to evaluate the vibration of elastically supported sections.

The two solver configurations therefore provide complementary aerodynamic information: fixed-body simulations characterize the loading and shedding behavior, while moving-body FSI simulations evaluate the motion-dependent response. The specific discretization schemes and coupling-iteration settings are summarized in Section~\ref{sec:3.3.1}, together with the mesh and time-step sensitivity assessments.

\subsection{Structural performance evaluation using OpenSees}
\label{sec:2.5}

OpenSees is an open-source finite-element framework for simulating the static and dynamic responses of structural systems. Its Python interface, OpenSeesPy, enables automated model construction, analysis, and response extraction within the proposed optimization framework \citep{Zhu2018}. For each candidate geometry, a section-analysis routine calculates the cross-sectional area \textit{A}, centroidal second moments of area \(I_{y}\) and \(I_{z}\), and Saint-Venant torsion constant \textit{J}. These properties are combined with the material parameters to define the axial, bending, and torsional stiffnesses, as well as the member mass distribution.

The structural model is assembled using three-dimensional elastic beam-column elements with prescribed boundary conditions and loadings. Static analysis provides internal forces and displacements while dynamic analysis determines natural frequencies and mode shapes. The resulting dynamic properties are transferred to the aerodynamic response model through the prescribed scaling relations. The sectional properties and computed internal forces are then used in the adopted strength and stability criteria to evaluate structural performance. In this way, the structural evaluation supplies both the performance measures used in optimization and the dynamic properties required for aerodynamic response analysis.

\subsection{Multi-objective Bayesian optimization}
\label{sec:2.6}

Multi-objective Bayesian optimization is adopted to explore the design space with a limited number of expensive numerical evaluations. Specifically, Gaussian-process (GP) surrogate models are combined with the qLogNParEGO acquisition function to select new design points. This approach is suited to CFD-FSI-based optimization because each candidate requires multiple transient simulations. The GP models provide predictions and uncertainty estimates, random scalarization explores different objective trade-offs, and batch selection enables parallel numerical evaluations \citep{Ament2023}.

Mathematically, the available observations at iteration \textit{t} are denoted by\(\mathcal{D}_{t}\). The design variables are normalized to the unit hypercube using their prescribed bounds:

\begin{equation}
\hat{x}_{r}=\frac{x_{r}-x_{r,\text{L}}}{x_{r,\text{U}}-x_{r,\text{L}}},\quad r=1,\ldots ,n.
\label{eq:19}
\end{equation}

For acquisition-function evaluation, all objectives are expressed in a common maximization direction, denoted by \(Y_{j}\). This is obtained by sign reversal \(Y_{j}=-f\), for the minimization formulation in Section~\ref{sec:2.1}. An independent GP is fitted to each objective using the normalized inputs, yielding the posterior prediction

\begin{equation}
Y_{j}(x)\mid \mathcal{D}_{t}\sim \mathcal{N}\left( \mu _{j,t}(x),\sigma _{j,t}^{2}(x)\right) ,\quad j=1,\ldots ,m.
\label{eq:20}
\end{equation}
where \(\mu _{j,t}\) and \(\sigma _{j,t}^{2}\) are the posterior mean and variance of objective \textit{j}, respectively. They describe the predicted performance and its uncertainty, providing the information needed to assess potential improvements before undertaking further numerical simulations.

The qLogNParEGO acquisition function combines random augmented Chebyshev scalarization with logarithmic noisy expected improvement. For each batch member, a weight vector is sampled from the unit simplex, \(\boldsymbol{w}\sim Dirichlet(1)\), with \(w_{j}>0\) and \(\sum_{j} w_{j}=1\). Following the BoTorch implementation, the maximization-oriented objectives are rescaled using the objective-wise ranges of the GP posterior means at the baseline designs. Denoting these rescaled responses by \(\widetilde{Y}_{j}\), the scalarized objective to be maximized is

\begin{equation}
g_{\boldsymbol{w}}(\widetilde{\boldsymbol{Y}})=-\left[ \operatorname*{max}\limits_{j=1,\ldots ,m}\left\{ w_{j}(1-\widetilde{Y}_{j})\right\} +\alpha _{s}\sum_{j=1}^{m} w_{j}(1-\widetilde{Y}_{j})\right]
\label{eq:21}
\end{equation}
where \(\alpha _{s}>0\) controls the augmentation term. Varying the weights directs the search towards different trade-off regions without prescribing a single fixed preference between objectives.

For a fixed weight vector, the unconstrained noisy expected improvement over a candidate set \(\mathcal{X}\) is expressed as

\begin{equation}
\begin{aligned}
\mathrm{NEI}_{\boldsymbol{w}}(\mathcal{X})
&= \mathbb{E}\Bigl[
  \Bigl( \max_{x\in\mathcal{X}} g_{\boldsymbol{w}}\bigl(\widetilde{\boldsymbol{Y}}(x)\bigr)\\
&\qquad {}-\max_{x\in x_{\mathrm{base}}}
  g_{\boldsymbol{w}}\bigl(\widetilde{\boldsymbol{Y}}(x)\bigr)
  \Bigr)_{+} \,\Bigm|\, \mathcal{D}_{t} \Bigr]
\end{aligned}
\label{eq:22}
\end{equation}
where \(x_{\text{base}}\) contains previously evaluated designs and \((z)_{+}=max(z,0)\). Candidate and baseline responses are sampled jointly from the GP posterior, so uncertainty in the best baseline response is included in the improvement calculation \citep{Bichon2013,BoTorch2025}. The logarithmic acquisition function can be represented conceptually as

\begin{equation}
\alpha _{\boldsymbol{w}}^{log}(\mathcal{X})\approx logNEI_{\boldsymbol{w}}(\mathcal{X}).
\label{eq:23}
\end{equation}

In practice, qLogNParEGO evaluates a numerically smooth approximation using Monte Carlo posterior samples. This formulation improves acquisition-function optimization when expected improvements become very small. Batch members are selected sequentially, with a new random scalarization for each candidate:

\begin{equation}
x_{t,b}\in \operatorname*{arg\,max}\limits_{x\in \mathcal{C}_{t,b}}\;\alpha _{\boldsymbol{w}_{b}}^{log}\left( x;\mathcal{D}_{t},X_{\text{pending}}\right) ,\;\;\;X_{\text{pending}}=\left\{ x_{t,1},\ldots ,x_{t,b-1}\right\} ,
\label{eq:24}
\end{equation}
where \textit{b} denotes the batch-member index and \(\mathcal{C}_{t,b}\) is the admissible candidate set. The completed batch is then submitted for parallel aerodynamic and structural analysis.

The adopted Bayesian optimizer provides one practical approach to computationally expensive multi-objective problems in structural wind engineering. The modular framework accommodates alternative surrogate models and acquisition functions, with the batch size adjusted to suit the design-space dimensionality, evaluation cost and engineering requirements of each application.

\subsection{Code Structure and Implementation}
\label{sec:2.7}

\begin{figure}[tbp]
\centering
\includegraphics[width=0.73\linewidth,height=0.73\textheight,keepaspectratio]{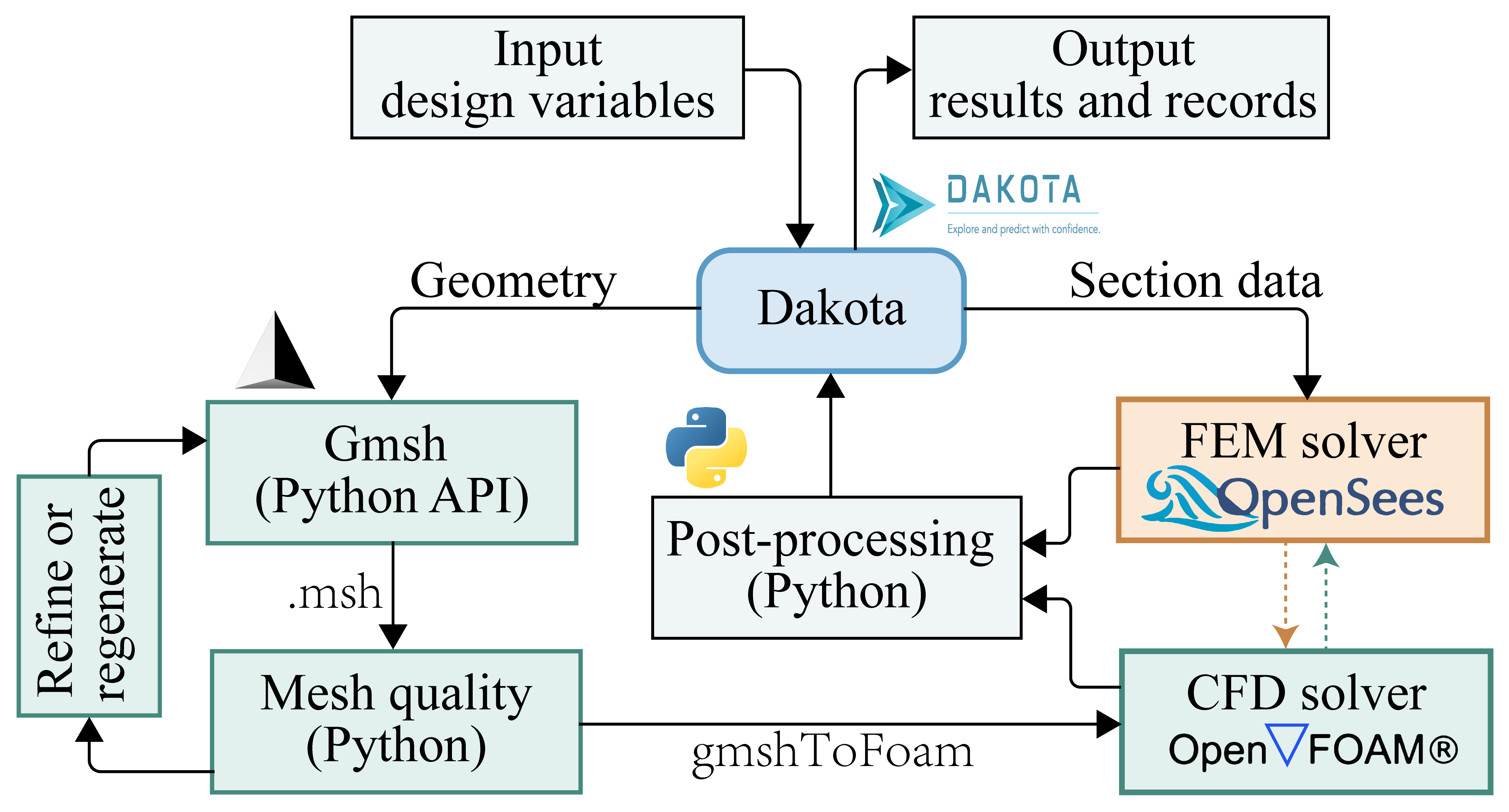}
\caption{Modular software architecture and data flow of the automated aero-structural optimization framework.}\label{fig:5}
\end{figure}

The implementation adopts a modular architecture connecting Dakota, Gmsh, OpenFOAM, and OpenSeesPy through Python drivers, as shown in Fig.~\ref{fig:5}. Dakota is an open-source toolkit developed by Sandia National Laboratories for design optimization, uncertainty quantification, and parameter estimation. It provides an extensible interface between simulation codes and iterative analysis methods \citep{Adams2023}. In the present framework, Dakota provides the design-variable and response interface, while the multi-objective Bayesian selection routine described in Section~\ref{sec:2.6} is implemented through Python.

For each candidate, the Python drivers pass the design variables to the geometry and structural modelling modules. The Gmsh Python API constructs the sectional geometry and generates the mesh, which is exported in \textit{.}\textit{msh} format and converted to OpenFOAM using \textit{gmshToFoam}. Mesh-quality checks trigger refinement or regeneration before flow analysis. The same variables update the sectional properties and the OpenSeesPy model. Structural dynamic parameters are transferred to the CFD-FSI module, while CFD-derived aerodynamic coefficients are supplied to the structural module to construct wind loads.

Python post-processing collects the aerodynamic and structural responses, and records the evaluation results. The results are returned through the Dakota interface and incorporated into the database used for subsequent surrogate updates and candidate selection. This modular organization allows individual analysis or optimization components to be modified without restructuring the entire workflow. The computational workflow was executed on a Linux server running Ubuntu 22.04.5 LTS, equipped with two AMD EPYC 7B13 processors providing 128 physical CPU cores in total.

\section{Problem formulation and numerical assessment}
\label{sec:3}

\begin{figure}[tbp]
\centering
\includegraphics[width=1.00\linewidth,height=0.73\textheight,keepaspectratio]{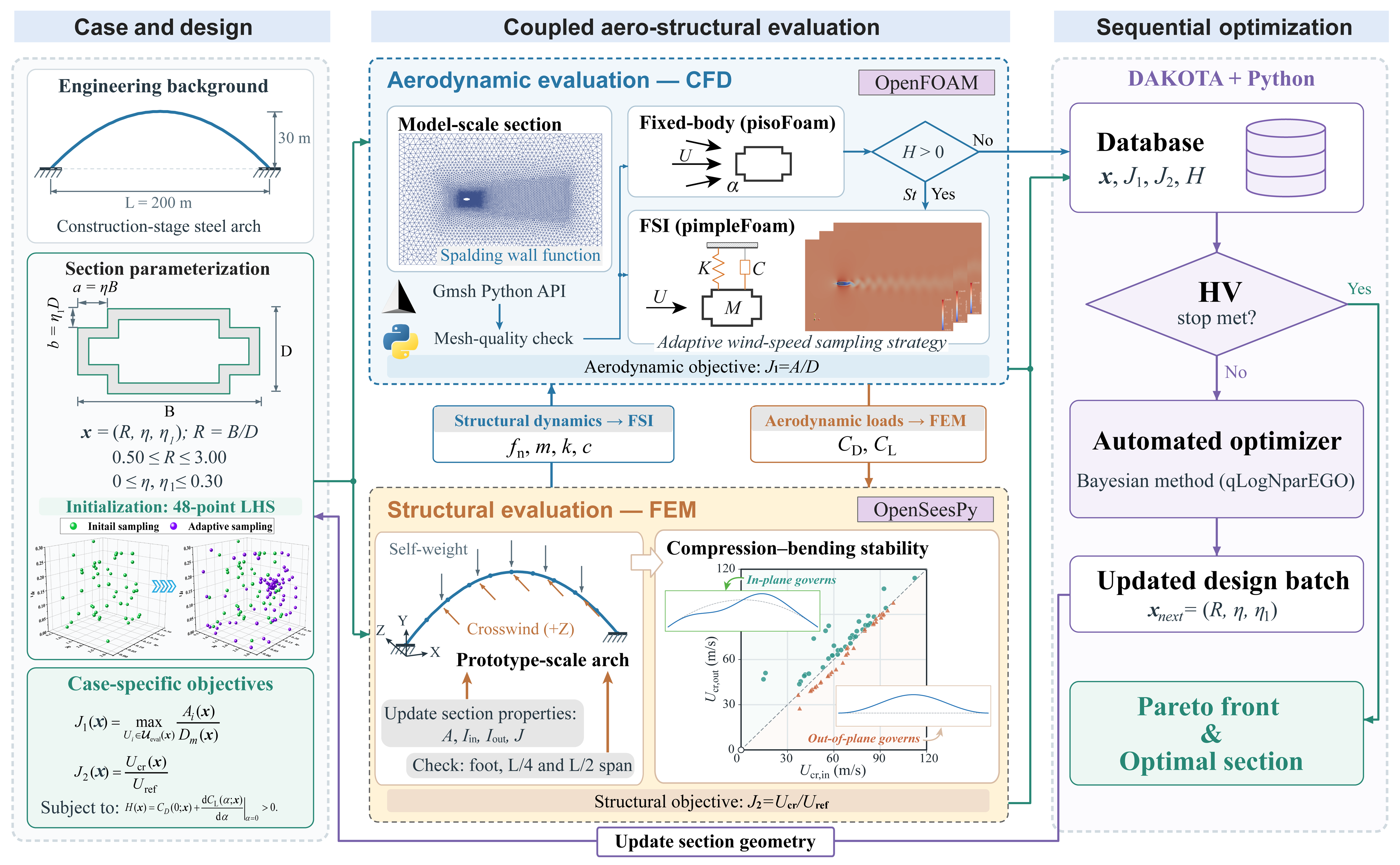}
\caption{Case-specific implementation of the aero-structural optimization framework for the construction-stage steel arch.}\label{fig:6}
\end{figure}

The framework presented in Section~\ref{sec:2} is applied to a construction-stage steel arch with parameterized corner-recessed rectangular sections. The case study examines how sectional geometry affects wind-induced vibration and structural stability, with normalized vibration amplitude and compression-bending stability margin serving as the competing objectives. Fig.~\ref{fig:6} summarizes the case-specific implementation, linking sectional design variables, coupled aerodynamic and structural evaluations, and sequential optimization. Sections~\ref{sec:3.1}-3.4 describe the engineering background, formulate the optimization problem, present the numerical models and verification studies, and specify the optimization settings.

\subsection{Description of the application case}
\label{sec:3.1}

\begin{figure}[tbp]
\centering
\includegraphics[width=0.87\linewidth,height=0.73\textheight,keepaspectratio]{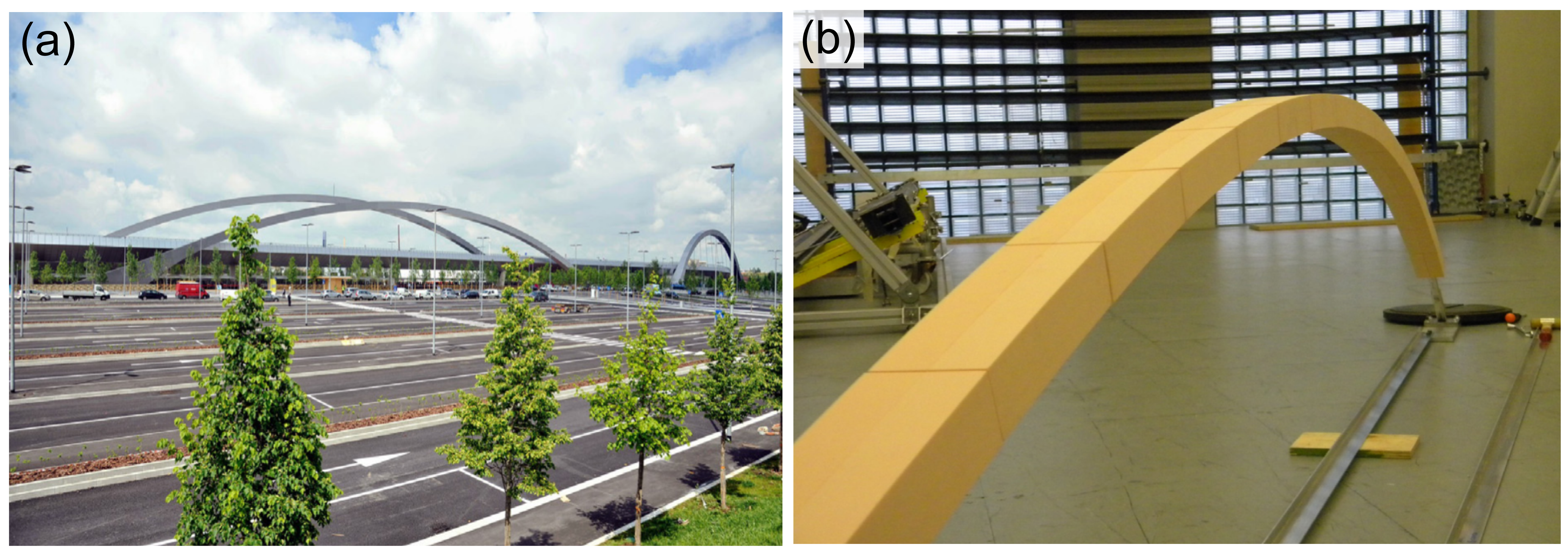}
\caption{Reference steel arch structures in Italy. \citep{Mannini2016}}\label{fig:7}
\end{figure}

The application case is inspired by the long-span steel arches constructed in Milan, Italy, for Expo 2015, as illustrated in Fig.~\ref{fig:7}. Designed as architectural landmarks rather than bridge-supporting structures, these arches combine slender structural forms with rectangular cross-sections, making wind-induced vibration an important design consideration. \citet{Mannini2016} investigated their aeroelastic behaviour through sectional and full-aeroelastic wind tunnel tests, highlighting the relevance of vortex-induced vibration and galloping to such structures.

Motivated by this engineering background, a steel arch is adopted to demonstrate the proposed optimization framework. The model represents a post-closure bare-arch construction condition, in which the arch is assessed under self-weight and transverse wind. The arch axis is parabolic, with a span of \textit{L} = 200 m and a rise of \textit{h} = 30 m, giving a rise-to-span ratio of 0.15. Both arch feet are fully fixed, restraining all three translational and three rotational degrees of freedom. The prototype-scale reference wind speed is set to 30 m/s.

The reference project supplies the engineering context, while the construction-stage and parameterized corner-recessed sections define the numerical application developed here. Aerodynamic interference from neighboring arches and surrounding buildings, as well as supplementary vibration-control devices, is excluded from the model. This setting isolates the influence of sectional geometry on wind-induced vibration and compression-bending stability, enabling consistent comparisons among candidate designs.

\subsection{Formulation of the aero-structural optimization problem}
\label{sec:3.2}

\subsubsection{Cross-sectional parameterization}
\label{sec:3.2.1}

\begin{figure}[tbp]
\centering
\includegraphics[width=0.60\linewidth,height=0.73\textheight,keepaspectratio]{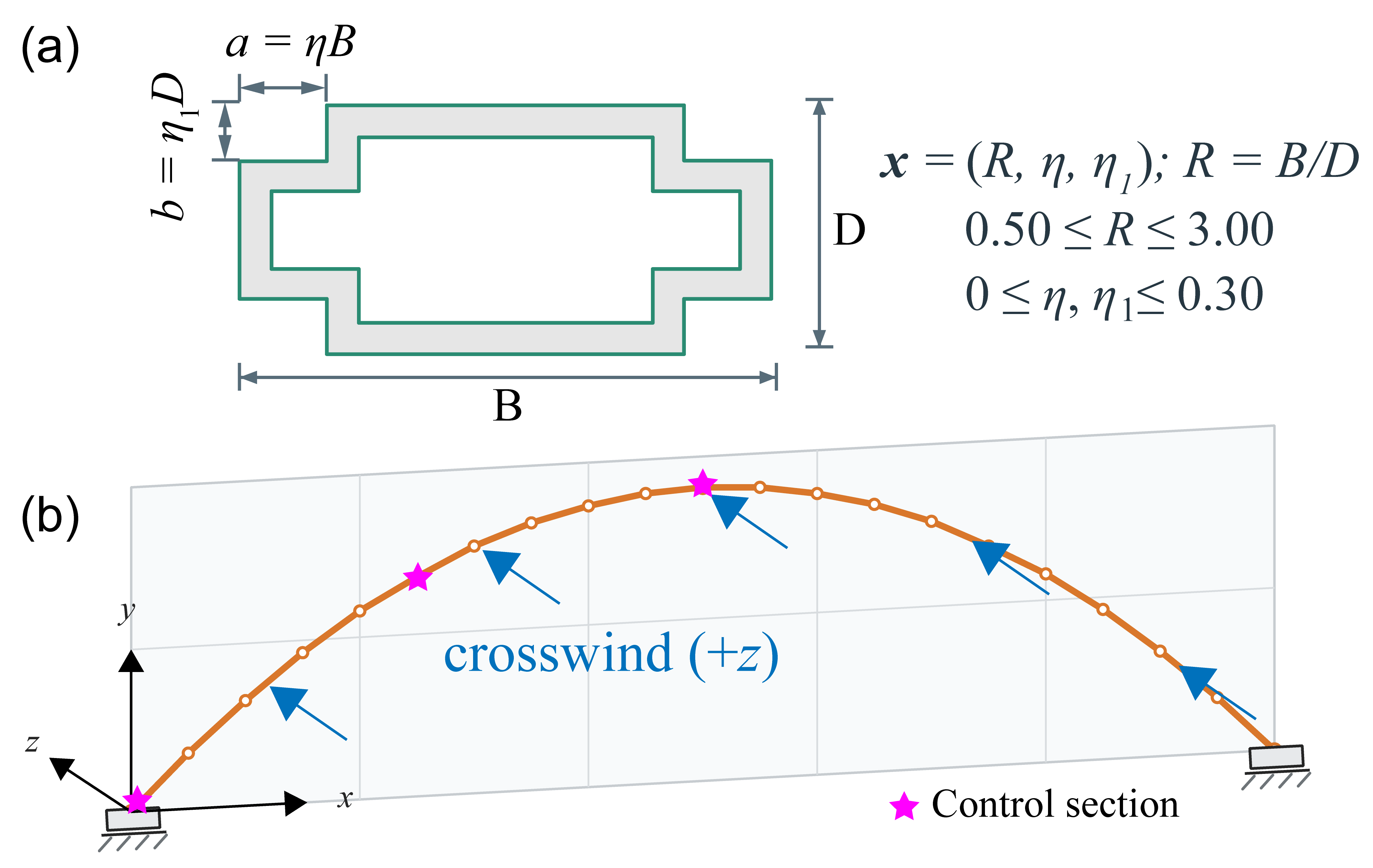}
\caption{Definition of the application model: (a) parameterization of the corner-recessed section; (b) control sections for structural stability assessment and transverse wind loading.}\label{fig:8}
\end{figure}

The optimization considers a doubly symmetric, closed steel section with rectangular recesses at its four corners, as illustrated in Fig.~\ref{fig:8}(a). The aerodynamic and structural models share the same design vector,

\begin{equation}
x=[R,\eta ,\eta _{1}]^{T},\quad R=\frac{B}{D},\quad a=\eta B,\quad b=\eta _{1}D.
\label{eq:25}
\end{equation}
where \textit{B} and \textit{D} are the overall sectional width and depth, respectively, and \textit{a} and \textit{b} specify the recess dimensions. The width is fixed at \textit{B} = 3.0 m, whereas the depth varies according to \textit{D} = \textit{B/R}. The design bounds are \(0.5\leq R\leq 3.0\) and \(0\leq \eta ,\eta _{1}\leq 0.30\). The wall thickness is \(t_{w}=26\;\text{mm}\). For the prescribed problem definition, the stress limit entering the structural interaction criteria is denoted by\(f_{d}=\sigma _{\text{allow}}=215\;\text{MPa}\).

For each candidate, the sectional area \(A_{\text{sec}}\), second moments of area \textit{I}\textsubscript{in} and \textit{I}\textsubscript{out}, torsion constant \textit{J}, and associated section moduli and radii of gyration provide the inputs to the finite-element model and compression-bending criteria. Detailed expressions are provided in Appendix~\ref{sec:A.1}.

\subsubsection{Aerodynamic objective}
\label{sec:3.2.2}

The aerodynamic objective minimizes the largest normalized equivalent vibration amplitude. For the vertical displacement history \(y_{i}\left( t\right)\) at wind speed \(U_{i}\), the mean displacement and equivalent amplitude over the selected statistical window \(\left[ t_{a},t_{b}\right]\) are defined as

\begin{equation}
\bar{y}_{i}=\frac{1}{T_{w}}\int_{t_{a}}^{t_{b}} y_{i}(t)\,\text{d}t,\quad A_{i}=\left[ \frac{2}{T_{w}}\int_{t_{a}}^{t_{b}} \left( y_{i}(t)-\bar{y}_{i}\right) ^{2}\,\text{d}t\right] ^{1/2},\quad T_{w}=t_{b}-t_{a}.
\label{eq:26}
\end{equation}

Using the sectional depth as the reference length, the aerodynamic objective is

\begin{equation}
J_{1}(x)=\operatorname*{max}\limits_{U_{i}\in \mathcal{U}_{\text{eval}}(x)}\frac{A_{i}(x)}{D_{m}(x)}
\label{eq:27}
\end{equation}
where \(\mathcal{U}_{\text{eval}}(x)\) is the set of successfully evaluated wind speeds generated by the adaptive procedure in Section~\ref{sec:2.2}. Thus, \(J_{1}\) characterizes the largest observed normalized response across the sampled wind speeds and provides a common vibration-performance measure for comparing candidate sections.

\subsubsection{Structural stability objective}
\label{sec:3.2.3}

The structural objective quantifies the wind-speed margin against compression-bending instability, accounting for sectional resistance, internal forces and aerodynamic loading. The lift coefficient and moment coefficients are expressed as

\begin{equation}
\begin{aligned}& \alpha _{\text{out}}=\frac{1}{2}\xi \rho DC_{\text{D}}L^{2}, \\ & \alpha _{\text{in}}=\frac{1}{2}\xi \rho BC_{\text{L}}L^{2}\end{aligned}
\label{eq:28}
\end{equation}
where the aerodynamic coefficients \(C_{\text{D}}\) and \(C_{\text{L}}\) are constructed from the static CFD analysis. The equivalent wind-moment coefficient is set to \(\xi =1/12\), based on the fixed-end moment of beam under a uniformly distributed transverse load. The resulting moments at wind speed \textit{U} are expressed as

\begin{equation}
\begin{aligned}& M_{\text{out},s}(U)=\alpha _{\text{out}}U^{2}, \\ & M_{\text{in},s}(U)=M_{G,s}-\alpha _{\text{in}}U^{2}\end{aligned}
\label{eq:29}
\end{equation}
where in-plane bending moment \(M_{G,s}\) refers to the maximum absolute values of the corresponding forces at the two element ends, which is extracted from FEM analysis. The structural assessment uses two coupled compression-bending interaction expressions. Their normalized utilization measures are \citep{Shen2005,MOHURD2017}

\begin{equation}
\begin{aligned}& r_{s,\text{in}}(U)=\frac{1}{f_{d}}[\frac{N_{s}}{\varphi _{\text{in}}A}+\frac{\beta _{m,\text{in}}M_{\text{in},s}(U)}{\gamma _{\text{in}}W_{\text{in}}\Delta _{s,\text{in}}}+\frac{\chi \beta _{t,\text{out}}M_{\text{out},s}(U)}{\varphi _{b}W_{\text{out}}}], \\ & r_{s,\text{out}}(U)=\frac{1}{f_{d}}[\frac{N_{s}}{\varphi _{\text{out}}A}+\frac{\beta _{m,\text{out}}M_{\text{out},s}(U)}{\gamma _{\text{out}}W_{\text{out}}\Delta _{s,\text{out}}}+\frac{\chi \beta _{t,\text{in}}M_{\text{in},s}(U)}{\varphi _{b}W_{\text{in}}}]\end{aligned}
\label{eq:30}
\end{equation}

with

\begin{equation}
\Delta _{s,d}=1-\frac{0.8N_{s}}{N_{E,d}^{\prime}},\;\;\;d\in \{\text{in},\text{out}\}.
\label{eq:31}
\end{equation}

Here, \(f_{d}\) is the stress limit specified in Section~\ref{sec:3.2.1}. \(\varphi _{d}\) and \(\varphi _{b}\) are the axial-compression and bending stability reduction factors, respectively; \(N_{E,d}^{\prime}\)is the reference elastic stability force; and \(\chi =0.7\) is the adopted closed-section influence coefficient. The equivalent-moment coefficients \(\beta _{m}\), \(\beta _{t}\) and the sectional coefficients \(\gamma _{d}\) are set to 1.0 in the present model. The supporting expressions and derivation of the compression-bending stability criteria are provided in Appendix~\ref{sec:A}.

For each control location \(\mathcal{S}\) and stability direction \textit{d}, the corresponding interaction condition is solved for the critical wind speed \(U_{s,d}(x)\) at which the stability limit is reached. The governing critical wind speed is the minimum over all control locations and both stability directions. The structural objective is defined as its ratio to the reference wind speed:

\begin{equation}
J_{2}(x)=\gamma(x)=\min_{\substack{s\in\mathcal{S}\\d\in\{\text{in},\text{out}\}}}
\frac{U_{s,d}(x)}{U_{\text{ref}}}
\label{eq:32}
\end{equation}
where \(U_{\text{ref}}=30\;\text{m/s}\). A larger \(\gamma\) indicates a greater stability, and structural feasibility \(\gamma \geq 1\). Three control locations are selected on one half of the arch---the arch foot, quarter-span, and crown---at

\begin{equation}
\mathcal{S}=\left\{ 0,\frac{L}{4},\frac{L}{2}\right\}
\label{eq:33}
\end{equation}

\subsubsection{Constrained multi-objective problem}
\label{sec:3.2.4}

A quasi-steady galloping constraint is imposed based on the Den Hartog criterion \citep{Macdonald2006}:

\begin{equation}
H(x)=C_{D}(0;x)+\left. \frac{\text{d}C_{L}(\alpha ;x)}{\text{d}\alpha}\right\rvert _{\alpha =0}>0.
\label{eq:34}
\end{equation}
\textit{H}<0 indicates susceptibility to classical transverse galloping through negative aerodynamic damping. This static screening criterion complements the transient vibration-response objective defined in Section~\ref{sec:3.2.2}.

Based on the aerodynamic and structural objectives defined in the preceding two subsections, the optimization problem is formulated as follows:

\begin{equation}
\begin{alignedat}{2}
&\operatorname*{minimize}_{x} &&\quad
\begin{bmatrix}J_{1}(x)\\-J_{2}(x)\end{bmatrix}\\
&\text{subject to} &&\quad x_{\text{L}}\leq x\leq x_{\text{U}}\\
&&&\quad H(x)>0\\
&&&\quad \gamma(x)\geq 1\\
&&&\quad x\in\mathcal{X}_{\text{geom}}
\end{alignedat}
\label{eq:35}
\end{equation}

The variable bounds and geometric admissibility conditions defining \(\mathcal{X}_{\text{geom}}\) are specified in Section~\ref{sec:3.2.1}.

\subsection{Numerical Setup}
\label{sec:3.3}

\subsubsection{CFD Setup}
\label{sec:3.3.1}

The aerodynamic evaluations are performed in OpenFOAM using a two-dimensional sectional model. The incompressible unsteady Reynolds-averaged Navier-Stokes equations are solved with the \textit{k}-\textit{\ensuremath{\omega}} SST turbulence model, following the formulation presented in Section~\ref{sec:2.4}. The air density and kinematic viscosity are set to \(\rho _{a}=1.225\;\text{kg/m}^{\text{3}}\) and \(\nu =1.5\times 10^{-5}\;\text{m}^{\text{2}}\text{/s}\), respectively. Fixed-body simulations provide the mean aerodynamic coefficients and vortex-shedding frequency, whereas elastically supported simulations determine the vibration response. The CFD sectional width is fixed at \(B_{m}=0.116\;\text{m}\), corresponding to a prototype width of \(B_{p}=3.0\;\text{m}\) and a geometric scale ratio of \(\lambda _{L}=26\). Under the prescribed velocity scaling, the upper assessment wind speed of 13.9 m/s at model scale corresponds to 30 m/s at prototype scale.

\begin{figure}[tbp]
\centering
\includegraphics[width=0.82\linewidth,height=0.73\textheight,keepaspectratio]{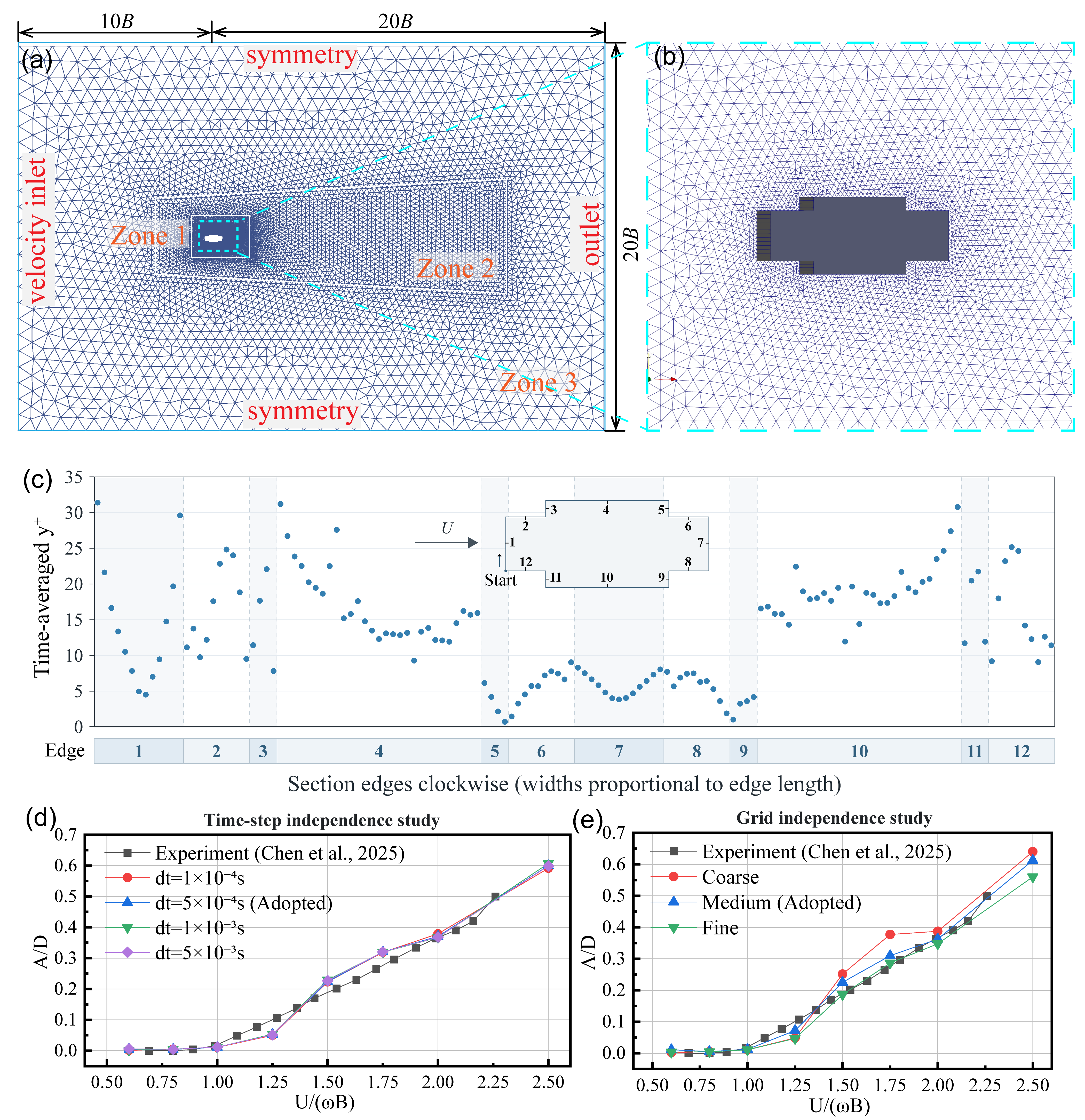}
\caption{CFD domain, mesh, and numerical sensitivity assessment: (a) domain dimensions and boundary conditions; (b) near-body mesh; (c) time-averaged \textit{y}\textsuperscript{+} distribution along the section; (d) time-step sensitivity; (e) mesh sensitivity.}\label{fig:9}
\end{figure}

The computational domain and representative mesh are illustrated in Fig.~\ref{fig:9}(a,b). Taking the sectional center as the origin, the domain extends from -10\textit{B} to 20\textit{B} in the streamwise direction and from -10\textit{B} to 10\textit{B} in the transverse direction. Its overall dimensions are therefore 30\textit{B} \ensuremath{\times} 20\textit{B}. A uniform mean velocity is prescribed at the inlet, together with a zero normal pressure gradient. At the outlet, the reference pressure is fixed and the velocity gradient is set to zero. Symmetry conditions are imposed on the upper and lower boundaries. The sectional surface satisfies a no-slip condition for fixed-body calculations and follows the body velocity during FSI calculations. The inlet turbulence intensity is \textit{I}=1\% and its length scale is \textit{L}=0.1\textit{B}.

The mesh is generated automatically using the Gmsh procedure described in Section~\ref{sec:2.3}. Three refinement regions are employed: Zone 1 resolves the sectional contour and corner recesses, Zone 2 extends downstream to resolve the wake, and Zone 3 provides a progressively coarser transition to the far-field boundaries. The planar triangular mesh is extruded through a single cell layer, with \textit{empty} conditions assigned to the front and back boundaries. Mesh validity, non-orthogonality, and skewness are checked after conversion to OpenFOAM. Near-wall turbulence is treated using \textit{nutUSpaldingWallFunction}\textit{,} as introduced in Section~\ref{sec:2.3}. Fig.~\ref{fig:9}(c) presents the time-averaged \textit{y}\textsuperscript{+} distribution along the section, with values ranging approximately from 1 to 32 for the illustrated case.

Fixed-body simulations are conducted using \textit{pisoFoam} at a depth-based Reynolds number of \textit{Re}\textsubscript{D}=1.5\ensuremath{\times}10\textsuperscript{4}. Aerodynamic coefficients are evaluated at incidence angles of -3\ensuremath{{}^{\circ}}, 0\ensuremath{{}^{\circ}} and 3\ensuremath{{}^{\circ}} for the galloping assessment, while the transient flow solution provides the vortex-shedding frequency. Moving-body simulations use \textit{pimpleFoam} with a deforming mesh. The sectional oscillator is restricted to vertical translation, with all other translational and rotational degrees of freedom constrained. Its mass, stiffness, and damping are updated for each candidate using the structural results and prescribed scaling relations, with a damping ratio of \(\zeta =0.16\%\). Body motion is integrated using the Newmark-\textit{\ensuremath{\beta}} scheme with \(\beta _{\text{N}}=0.25\) and \(\gamma _{\text{N}}=0.5\). The mesh deforms around the moving section while the far-field boundaries remain fixed. Across the sampled wind speeds and candidate geometries, the FSI Reynolds number ranges approximately from 1.4\ensuremath{\times}10\textsuperscript{3} to 2.3\ensuremath{\times}10\textsuperscript{5}.

The solver settings and discretization schemes for the fixed-body and moving-mesh simulations are summarized in Table~\ref{tab:1}. Both solvers employ second-order implicit backward time discretization and linearupwind discretization for momentum convection, whereas the turbulence convection terms use linearupwind discretization. Pressure is solved using a geometric-algebraic multigrid method, while velocity and turbulence variables use smooth iterative solvers.

\begin{table}[tbp]
\centering
\caption{Solver settings and discretization schemes for the OpenFOAM simulations.}\label{tab:1}
\small
\begin{tabularx}{\linewidth}{@{}p{0.27\linewidth}LL@{}}
\toprule
Item & Fixed-body analysis & FSI analysis \\
\midrule
OpenFOAM solver & pisoFoam & pimpleFoam \\
Pressure-velocity coupling & PISO & PIMPLE \\
Turbulence model & \textit{k-\ensuremath{\omega}} SST & \textit{k-\ensuremath{\omega}} SST \\
Time discretization & backward & backward \\
Gradient & leastSquares & leastSquares \\
Momentum convection & Gauss linearUpwind grad(U) & Gauss linearUpwind grad(U) \\
\textit{K, \ensuremath{\omega}} convection & Gauss linearUpwind & Gauss linearUpwind \\
Laplacian & Gauss linear corrected & Gauss linear corrected \\
Face interpolation & linear & linear \\
Surface-normal gradient & corrected & corrected \\
Pressure linear solver & GAMG\textsuperscript{\textit{a}} & GAMG\textsuperscript{\textit{a}} \\
\textit{U, k, \ensuremath{\omega}} linear solver & smoothSolver; symGauss\allowbreak Seidel & smoothSolver; symGauss\allowbreak Seidel \\
Outer corrections & - & \textit{nOuterCorrectors} = 10 \\
Pressure-velocity corrections & \textit{nCorrectors} = 3 & nCorrectors = 2 \\
Non-orthogonal corrections & \textit{nNonOrthogonal\allowbreak Correctors}=1 & \textit{nNonOrthogonal\allowbreak Correctors}= 2 \\
Near-wall treatment & \textit{nutUSpalding\allowbreak WallFunction} & \textit{nutUSpalding\allowbreak WallFunction} \\
Time-step & 5\ensuremath{\times}10\textsuperscript{-4} s & 5\ensuremath{\times}10\textsuperscript{-4} s \\
\bottomrule
\end{tabularx}
\vspace{2pt}
\begin{minipage}{\linewidth}
\raggedright\footnotesize
\textsuperscript{\textit{a}} \textit{GAMG: Geometric-Algebraic MultiGrid method}
\end{minipage}
\end{table}

Time-step and mesh sensitivity are examined using the reference vibration-response case shown in Fig.~\ref{fig:9}(d,e), with the experimental measurements of \citet{Chen2025} included for comparison. The time-step curves are closely grouped over most of the tested velocity interval and reproduce the overall experimental response trend. The coarse mesh overpredicts the vibration amplitude over part of the interval, whereas the medium and fine meshes generally provide closer agreement with the measurements. Accordingly, the medium mesh and dt = 5\ensuremath{\times}10\textsuperscript{-4} s are adopted for the optimization evaluations.

\subsubsection{Structural finite-element setup}
\label{sec:3.3.2}

\begin{figure}[tbp]
\centering
\includegraphics[width=0.78\linewidth,height=0.73\textheight,keepaspectratio]{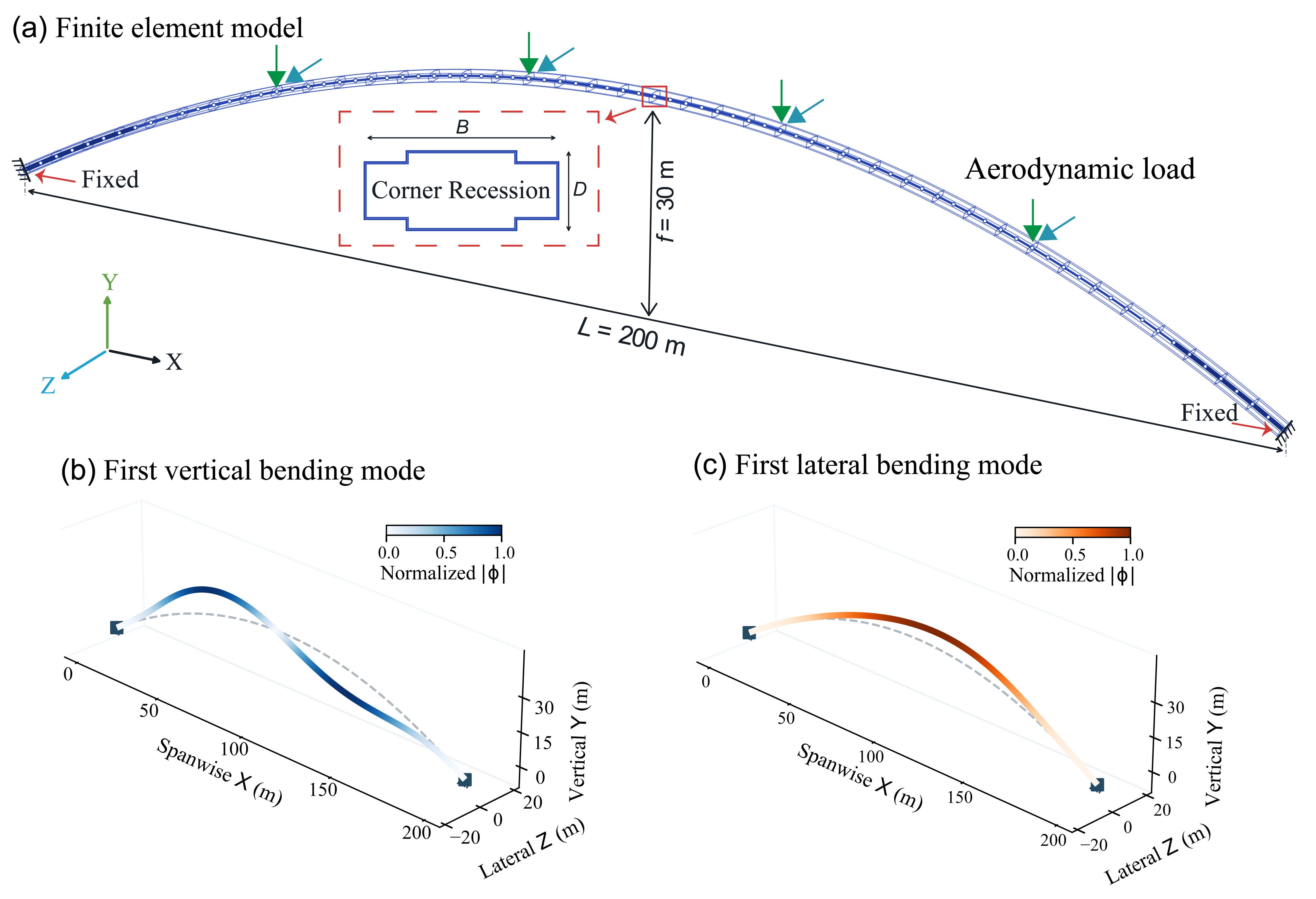}
\caption{Structural finite-element model and representative bending modes: (a) three-dimensional arch finite-element model; (b) first in-plane vertical bending mode; (c) first out-of-plane lateral bending mode.}\label{fig:10}
\end{figure}

The structural quantities required by Section~\ref{sec:3.2} are obtained using a three-dimensional beam model implemented in OpenSeesPy, as illustrated in Fig.~\ref{fig:10}(a). The arch geometry follows Section~\ref{sec:3.1}, and each node has three translational and three rotational degrees of freedom. The global \textit{X}, \textit{Y}, and \textit{Z} axes denote the horizontal spanwise, vertical, and transverse directions, respectively. For each element, the local \textit{x}-axis follows the element centerline, the local \textit{z}-axis is aligned with the global \textit{Z}-axis, and the local \textit{y}-axis lies in the arch plane. Accordingly, \(I_{z}=I_{\text{in}}\) governs in-plane bending and \(I_{y}=I_{\text{out}}\) governs out-of-plane bending.

Steel is modelled as a linear-elastic material with Young's modulus\(E=2.06\times 10^{11}\;\text{Pa}\), shear modulus \(G=7.90\times 10^{10}\;\text{Pa}\), and density \(\rho _{s}=7850\;\text{kg/m}^{\text{3}}\). The wall thickness is 26 mm, and the sectional properties are updated for each design geometry. Both arch feet are fully restrained against translation and rotation. The loading represents the post-closure bare-arch condition without bridge-deck or traffic loads. Static deformation is assessed under combined gravity and wind loading. For stability assessment, wind effects are incorporated separately through the expressions in Section~\ref{sec:3.2.3}.

Modal analysis extracts the first eight elastic vibration modes using the linear stiffness and mass matrices. Modes are classified by their dominant nodal translation directions and verified through mode-shape inspection. The lowest-frequency modes dominated by vertical and transverse displacement are identified as the first in-plane vertical and out-of-plane lateral bending modes, respectively, as illustrated in Fig.~\ref{fig:10}(b,c). It should be noted that this approach accommodates changes in modal ordering as the sectional geometry varies, rather than relying on fixed mode numbers. The modal predictions are checked against the full-scale calculated frequencies reported by \citet{Mannini2016}, using the reference rectangular section with \textit{R}=1.5. Table~\ref{tab:2} compares the four principal bending frequencies with the published full-scale calculated frequencies. The calculated low-order bending frequencies agree closely with the reference values, with a maximum relative deviation of 2.84\%.

\begin{table}[tbp]
\centering
\caption{Verification of the OpenSeesPy model through comparison of full-scale bending frequencies. (OP: out of plane; IP: in plane; BS: bending symmetric; BA: bending antisymmetric)}\label{tab:2}
\small
\begin{tabularx}{\linewidth}{@{}lCCC@{}}
\toprule
Mode & Reference value (Hz) \citep{Mannini2016} & Present model by  Openseespy (Hz) & Relative error \\
\midrule
OP-BS & 0.55 & 0.559 & +1.63 \% \\
IP-BA & 1.17 & 1.197 & +2.30 \% \\
OP-BA & 1.48 & 1.522 & +2.84 \% \\
IP-BS & 1.94 & 1.945 & +0.25 \% \\
\bottomrule
\end{tabularx}
\end{table}

\subsection{Optimization Setup}
\label{sec:3.4}

The optimization is organized into two stages: initial design-space sampling and sequential adaptive enrichment. In Stage I, an initial set of 48 designs is obtained through Latin hypercube sampling (LHS) within the three-variable design domain defined in Section~\ref{sec:3.2}, as shown in Fig.~\ref{fig:11}(a). Each design is evaluated using the coupled aerodynamic-structural procedure to obtain the vibration-response objective, structural stability indicator. These evaluations establish the initial database for subsequent surrogate-assisted exploration.

In Stage II, the database is enriched in batches of five designs. Each batch contains three Pareto-improvement candidates selected using constrained qLogNParEGO, one uncertainty-driven exploration candidate, and one maximin exploration candidate. This allocation combines improvement-oriented sampling with exploration of uncertain and sparsely sampled regions. At each iteration, a pool of 3000 uniformly sampled designs is generated and filtered. The selected designs undergo direct numerical evaluation, including the adaptive wind-speed sampling procedure in Section~\ref{sec:2.2}. Successful evaluations are added to the database, and the Gaussian-process models are refitted before the next batch is selected.

Optimization progress is monitored using the hypervolume of the evaluated set. The relative HV improvement at iteration \textit{t} is calculated as

\begin{equation}
\delta _{\text{HV}}^{(t)}=\frac{\text{HV}_{t}-\text{HV}_{t-1}}{\text{HV}_{t-1}},\quad \text{HV}_{t-1}>0.
\label{eq:36}
\end{equation}

The optimization is considered converged when the hypervolume shows no improvement for five consecutive iterations. The detailed settings are summarized in Table~\ref{tab:3}.

\begin{table}[tbp]
\centering
\caption{Optimization settings of the case.}\label{tab:3}
\small
\begin{tabularx}{\linewidth}{@{}p{0.40\linewidth}L@{}}
\toprule
Item & Setting \\
\midrule
Initial sampling & 48 Latin hypercube samples \\
Surrogate models & Gaussian-process models \\
Input normalization & Design variables mapped to [0,1]\textsuperscript{3} \\
GP hyperparameter fitting & Marginal-likelihood maximization \\
Pareto-improvement acquisition & Constrained qLogNParEGO \\
Scalarization weights & Independent \(w\sim\operatorname{Dirichlet}(1,1)\) \\
Posterior sampling & 128 Sobol quasi-Monte Carlo samples \\
Candidate pool & 3000 uniformly sampled designs per iteration \\
Stage II batch size & 5 designs \\
Minimum sample separation & 0.025 in normalized Euclidean distance \\
Exclusion radius around registered forbidden points & 0.04 in normalized design space \\
\bottomrule
\end{tabularx}
\end{table}

\section{Optimization results and discussion}
\label{sec:4}

Following the optimization settings in Section~\ref{sec:3}, this section examines the sampling evolution and observed Pareto front. Representative designs are subsequently compared in terms of their geometry, aerodynamic response, and structural performance, followed by a discussion of engineering trade-offs and design selection.

\subsection{Sampling evolution and Pareto Front}
\label{sec:4.1}

\begin{figure}[tbp]
\centering
\includegraphics[width=0.82\linewidth,height=0.62\textheight,keepaspectratio]{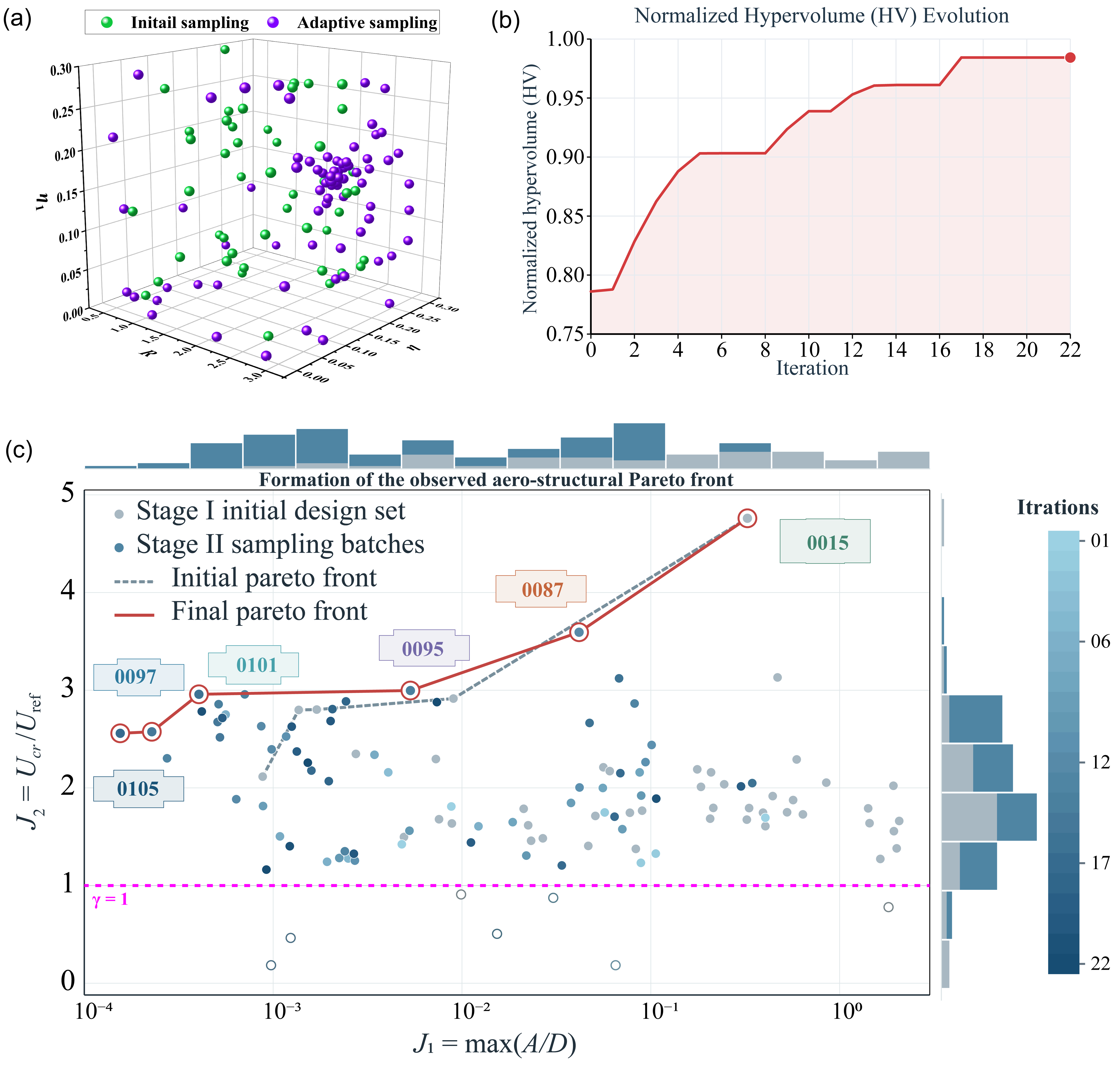}
\caption{Sampling evolution and observed aero-structural trade-offs: (a) design-space distribution of Stage I and Stage II samples; (b) HV evolution; (c) initial and final observed Pareto fronts with marginal objective distributions and the six labelled representative designs.}\label{fig:11}
\end{figure}

Fig.~\ref{fig:11} summarizes the sampling evolution and the resulting aero-structural trade-offs. The retained analysis dataset comprises 123 designs, including 48 initial designs in Stage I and 75 additional designs distributed across 22 sequential batches in Stage II. A leave-one-out cross-validation (LOOCV) assessment of reconstructed GP surrogates is presented in Appendix~\ref{sec:B}. The three-dimensional distribution in Fig.~\ref{fig:11}(a) shows broad initial coverage followed by more concentrated sampling during Stage II. The proportion of designs with 2< R<3 increases from 41.7\% in Stage I to 81.3\% in Stage II. As shown in Fig.~\ref{fig:11}(b), the reconstructed HV increases from 0.7861 to 0.9843, corresponding to a relative increase of 25.2\%. After batch 17, another 16 designs are evaluated without changing the retained observed front or its HV.

Fig.~\ref{fig:11}(c) compares the initial and final Pareto fronts. Sequential sampling improves the observed trade-off set, shifting the initial front, shown in grey, toward the upper-left region of the objective space to form the final red front. The final front contains six non-dominated designs, five obtained during Stage II and one retained from Stage I, namely design 0015. Along this front, increasing the structural indicator is accompanied by a larger sampled vibration-response envelope. Design 0105 provides the lowest observed aerodynamic objective, \textit{J}\textsubscript{1}=1.5501\ensuremath{\times}10\textsuperscript{-4}, with \textit{J}\textsubscript{2}=2.5589, whereas design 0015 provides the largest finite structural indicator, \textit{J}\textsubscript{2}=4.7603, with \textit{J}\textsubscript{1}=0.3247. Their aerodynamic and structural performance are examined in the following subsections, and preference-based selection is discussed in Section~\ref{sec:4.5}.

\subsection{Effects of design variables on the objectives}
\label{sec:4.2}

\begin{figure}[tbp]
\centering
\includegraphics[width=0.74\linewidth,height=0.62\textheight,keepaspectratio]{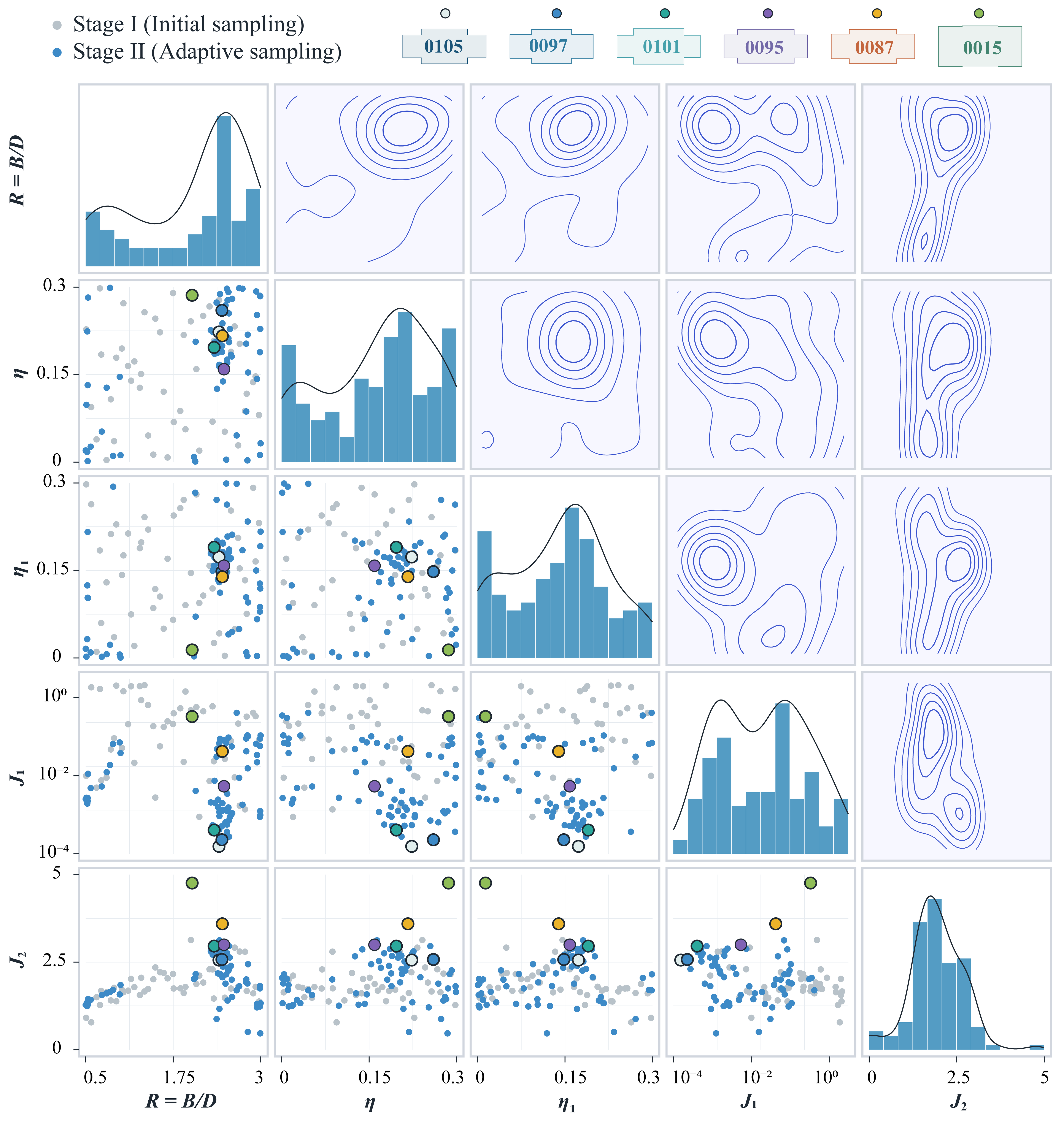}
\caption{Associations among the design variables and evaluated responses.}\label{fig:12}
\end{figure}

The relationships between sectional geometry and optimization performance are examined using pairwise sample distributions, surrogate response surfaces, and SHAP analysis, as presented in Figs.~\ref{fig:12}-14. Together, these analyses reveal the parameter combinations associated with favorable responses and the relative contributions of the three design variables to the predicted objectives.

Fig.~\ref{fig:12} summarizes the pairwise associations among the three design variables and the two objectives. The lower triangle distinguishes the evaluated designs by sampling stage, with outlined markers identifying the six observed Pareto designs. The diagonal shows marginal distributions, while the upper triangle presents bivariate sampling-density contours. The design-variable panels show that Stage I provides broad range, whereas Stage II concentrates more strongly around \textit{R} \ensuremath{\approx} 2.3-2.5 and intermediate corner-recess dimensions. Five of the six Pareto designs lie within this region. Nevertheless, the corresponding \textit{J}\textsubscript{1}-\textit{R} and \textit{J}\textsubscript{2}-\textit{R} panels show substantial response variation at similar aspect ratios, indicating that \textit{R} alone cannot determine performance. The corner-recession parameters therefore remain important for distinguishing designs with similar overall proportions. The response panels further reveal the competing objectives. Several Pareto designs achieve very low sampled vibration responses, whereas the design with the highest structural stability margin exhibits a larger aerodynamic response. Their positions in the \textit{J}\textsubscript{1}-\textit{J}\textsubscript{2} panel illustrate the resulting trade-off. The marginal distributions also highlight the wide range of aerodynamic responses and the relatively small number of designs attaining high structural margins.

\begin{figure}[!p]
\centering
\includegraphics[width=0.76\linewidth,height=0.54\textheight,keepaspectratio]{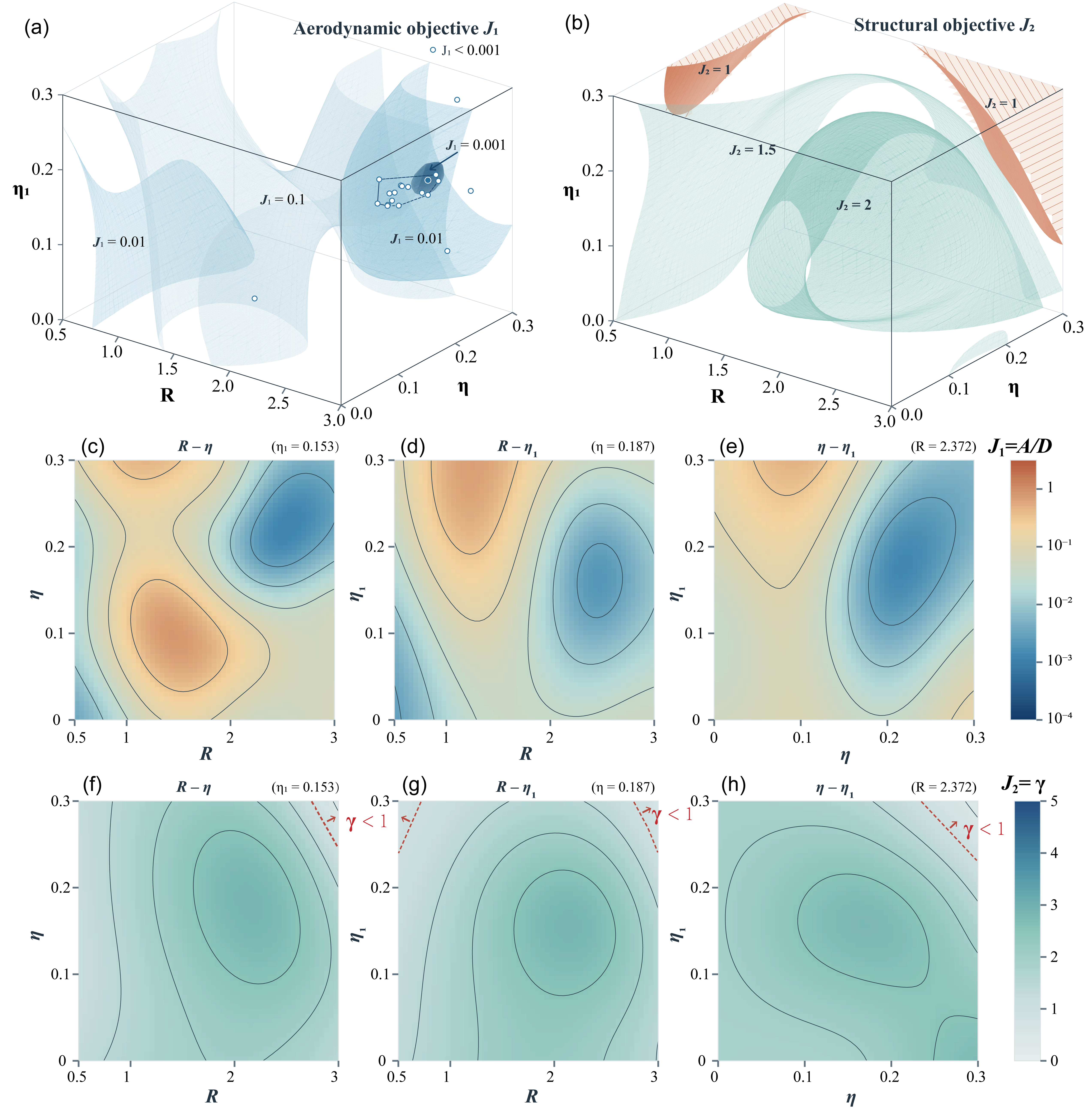}
\caption{Reconstructed GP response surfaces: (a,b) three-dimensional isosurfaces of \textit{J}\textsubscript{1} and \textit{J}\textsubscript{2}; (c-e) aerodynamic response slices in the \textit{R}--\textit{\ensuremath{\eta}}, \textit{R}--\textit{\ensuremath{\eta}}\textsubscript{1}, and \textit{\ensuremath{\eta}}--\textit{\ensuremath{\eta}}\textsubscript{1} planes; (f-h) structural response slices in the \textit{R}--\textit{\ensuremath{\eta}}, \textit{R}--\textit{\ensuremath{\eta}}\textsubscript{1}, and \textit{\ensuremath{\eta}}--\textit{\ensuremath{\eta}}\textsubscript{1} planes.}\label{fig:13}
\end{figure}

Fig.~\ref{fig:13} further characterizes these relationships using reconstructed GP model. The three-dimensional isosurfaces illustrate selected response levels, and the two-dimensional slices provide more readable comparisons between variable pairs. In each slice, the omitted variable is fixed at its sample median: \textit{R}=2.372, \textit{\ensuremath{\eta}}=0.187 or \textit{\ensuremath{\eta}}\textsubscript{1}=0.153.

The aerodynamic contours contain localized low-response regions rather than a uniform decrease toward a single design-domain boundary. In the higher-\textit{R} region, the predicted response depends strongly on the combination of the two corner-recess dimensions, with a low-response basin occurring at intermediate \textit{\ensuremath{\eta}}\textsubscript{1} and moderate-to-large \textit{\ensuremath{\eta}} in the displayed slices. The structural contours instead show a broader region of elevated \textit{J}\textsubscript{2}, with reductions toward some parameter-domain boundaries. The favorable regions of the two objectives overlap but do not coincide, providing a model-based illustration of the trade-off observed in Section~\ref{sec:4.1}. These patterns also indicate that neither increasing \textit{R} nor enlarging both corner-recessed ratios produces the improvement in performance.

\begin{figure}[!p]
\centering
\includegraphics[width=0.66\linewidth,height=0.40\textheight,keepaspectratio]{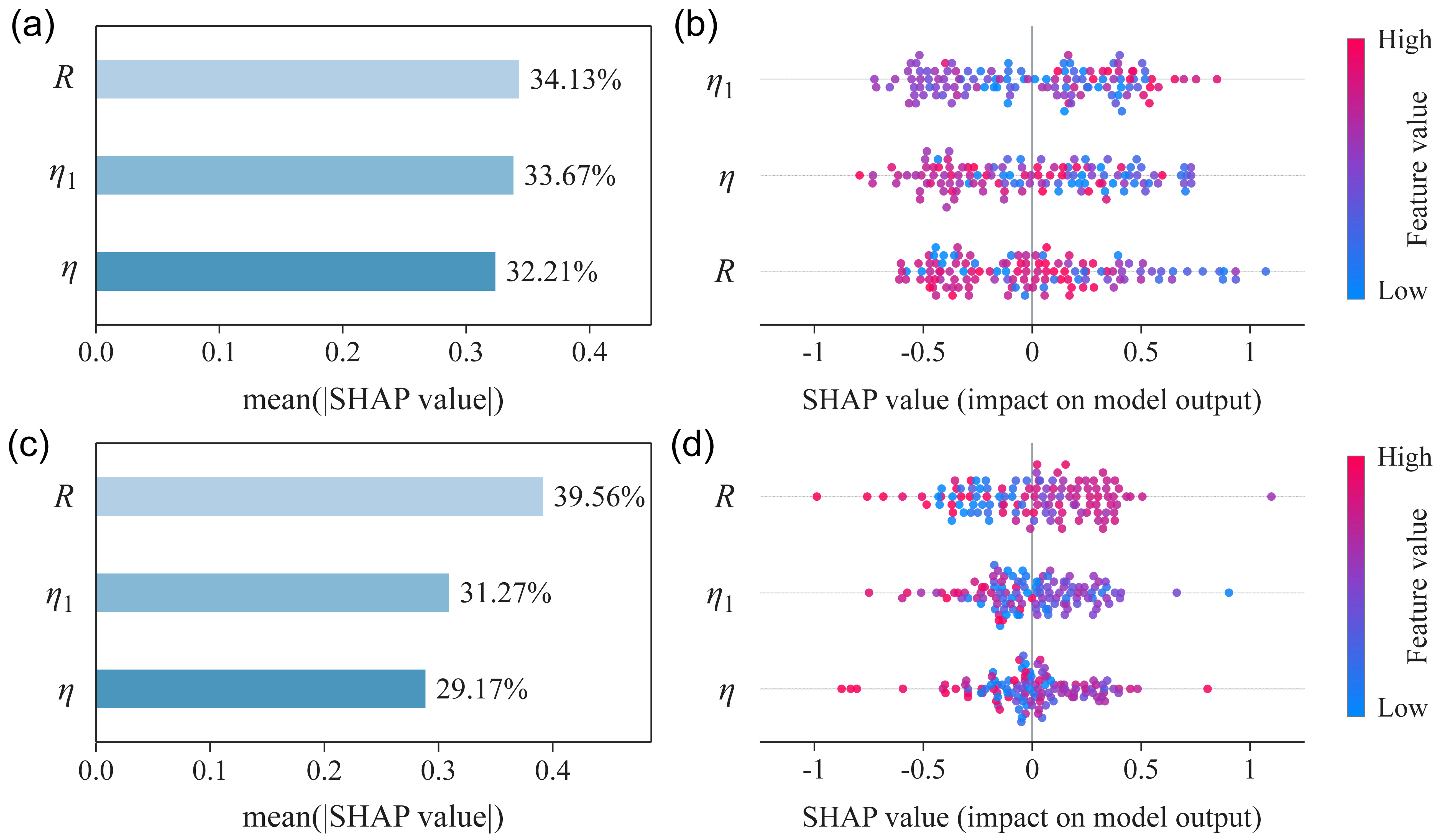}
\caption{SHAP analysis of the aerodynamic and structural surrogate models: (a) importance ranking of design variables for aerodynamic objective; (b) summary plot of design variables for aerodynamic objective; (c) importance ranking of design variables for structural objective; (d) summary plot of design variables for structural objective.}\label{fig:14}
\end{figure}

SHapley Additive exPlanations (SHAP) are used to quantify the contributions of the design variables to the optimization objectives, as shown in Fig.~\ref{fig:14}. The mean absolute SHAP values summarize overall variable importance, while the beeswarm plots show the magnitude and direction of individual contributions. Point colours indicate the corresponding input values. Negative contributions to \textit{J}\textsubscript{1} favour vibration mitigation, whereas positive contributions to \textit{J}\textsubscript{2} favour structural stability.

For the aerodynamic objective, \textit{R}, \textit{\ensuremath{\eta}}\textsubscript{1}, and \textit{\ensuremath{\eta}} have comparable importance, accounting for 34.1\%, 33.7\%, and 32.2\% of the total mean absolute SHAP value, respectively. For the structural objective, \textit{R} is the most influential individual variable, while the two corner-recession parameters together account for 60.4\% of the total attribution. Thus, the aerodynamic response depends on all three variables to a similar extent, whereas structural performance exhibits a stronger individual dependence on the overall sectional proportion while remaining sensitive to corner modifications. Combined with the response surfaces, these results support joint optimization of \textit{R}, \textit{\ensuremath{\eta}}, and \textit{\ensuremath{\eta}}\textsubscript{1}.

\subsection{Aerodynamic performance}
\label{sec:4.3}

\subsubsection{Galloping performance}
\label{sec:4.3.1}

Rectangular sections can experience large-amplitude galloping and interactions with vortex-induced vibration, depending on sectional geometry and the mass--damping characteristics of the supporting system \citep{Mannini2014,Gao2017}. Here, galloping susceptibility is evaluated using the quasi-steady Den Hartog indicator \textit{H}, defined in Eq.~\eqref{eq:34} of Section~\ref{sec:3.2.4}.

\begin{figure}[tbp]
\centering
\includegraphics[width=0.98\linewidth,height=0.73\textheight,keepaspectratio]{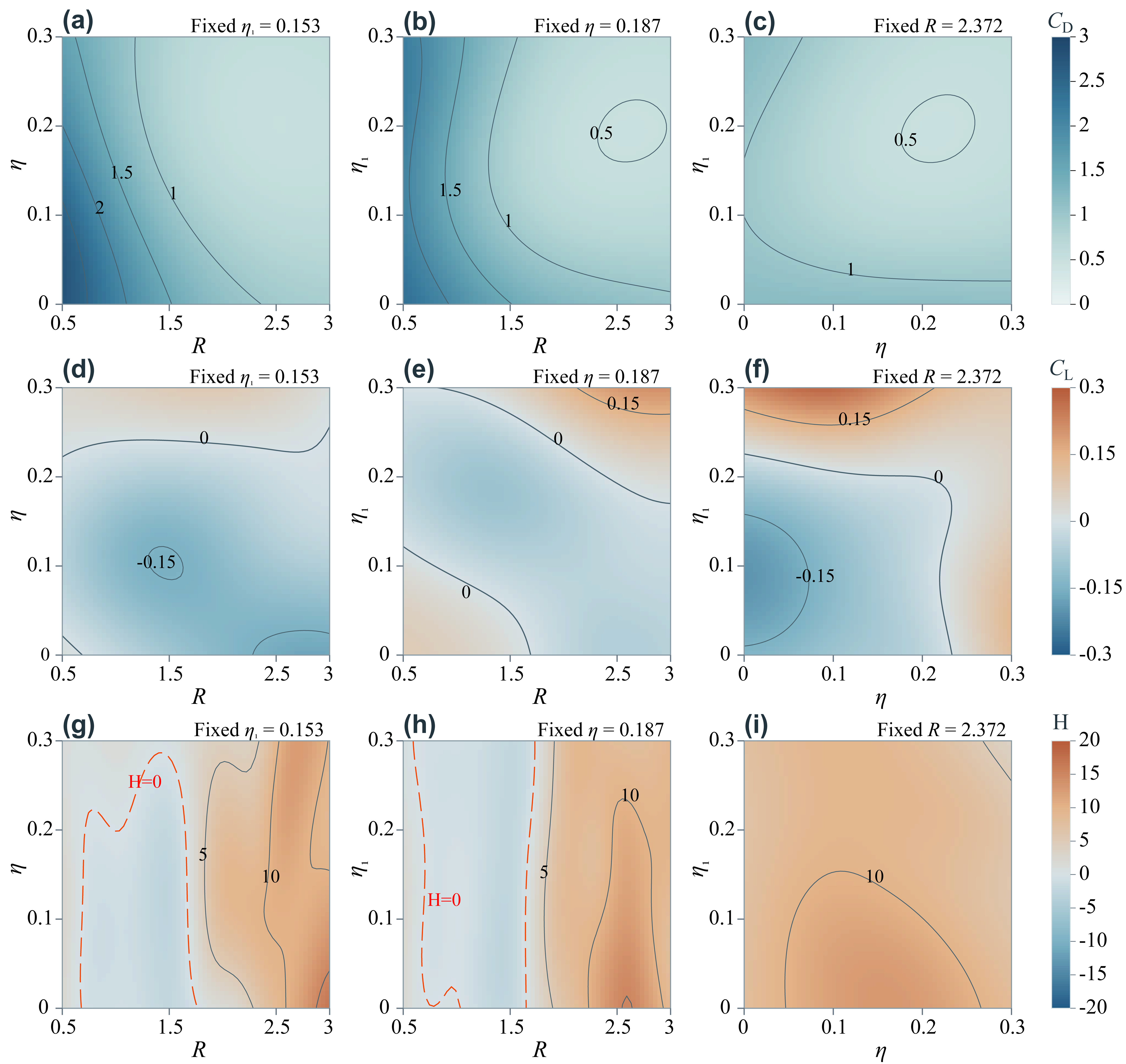}
\caption{Static-aerodynamic response contours: (a--c) drag coefficient \(C_{D}\); (d--f) lift coefficient \(C_{L}\); and (g--i) Den Hartog indicator \textit{H}.}\label{fig:15}
\end{figure}

Fig.~\ref{fig:15} presents reconstructed contours of \(C_{D}\), \(C_{L}\), and \textit{H} based on fixed-body evaluations at \textit{Re}=15000 and incidence angles of -3\ensuremath{{}^{\circ}}, 0\ensuremath{{}^{\circ}}, and +3\ensuremath{{}^{\circ}}. Figs.~\ref{fig:15}(a-c) show that increasing \textit{R} generally reduces the predicted drag coefficient, although the corner recession modify this trend. At fixed \textit{\ensuremath{\eta}}\textsubscript{1} = 0.153, increasing \textit{\ensuremath{\eta}} generally reduces drag in the lower-\textit{R} region, as indicated by the inclined constant-drag contours in Fig.~\ref{fig:15}(a). The effect of \textit{\ensuremath{\eta}}\textsubscript{1} is non-monotonic: Fig.~\ref{fig:15}(b) contains a closed \(C_{D}=0.5\) contour at higher \textit{R} and intermediate \textit{\ensuremath{\eta}}\textsubscript{1}. Similarly, Fig.~\ref{fig:15}(c) shows a localized low-drag region for intermediate combinations of \textit{\ensuremath{\eta}} and \textit{\ensuremath{\eta}}\textsubscript{1}. Enlarging both recesses therefore does not produce a continuous reduction in drag.

The displayed lift-coefficient contours show changes in both magnitude and sign. In Fig.~\ref{fig:15}(d), negative \(C_{L}\) occupies much of the lower-to-intermediate-\textit{\ensuremath{\eta}} region, while increasing \textit{\ensuremath{\eta}} toward its upper bound shifts the mean lift toward zero or positive values. At \textit{R}=2.372, Fig.~\ref{fig:15}(f) shows a similar shift with increasing \textit{\ensuremath{\eta}}\textsubscript{1} or as \textit{\ensuremath{\eta}} approaches its upper bound.

Figs.~\ref{fig:15}(g,h) show localized negative-\textit{H} regions at lower-to-intermediate \textit{R}, bounded by curved \textit{H} = 0 contours. Their positions vary with the corner-recession ratios, demonstrating that no single \textit{R} threshold separates favorable and unfavorable combinations throughout the domain. Higher-\textit{R} combinations generally exhibit positive \textit{H}. In particular, no \textit{H} = 0 boundary appears in Fig.~\ref{fig:15}(i), and the reconstructed response remains positive over the displayed plane at \textit{R} = 2.372. Nevertheless, the magnitude of \textit{H} still varies: moderate \textit{\ensuremath{\eta}} combined with lower-to-intermediate \textit{\ensuremath{\eta}}\textsubscript{1} gives higher values than the upper-right region with both corner-recession ratios large. The coexistence of lower drag and positive \textit{H} also shows why drag reduction alone cannot explain the galloping trend---the lift-slope contribution should be considered.

\subsubsection{Vortex-induced vibration performance}
\label{sec:4.3.2}

\begin{figure}[tbp]
\centering
\includegraphics[width=1.00\linewidth,height=0.73\textheight,keepaspectratio]{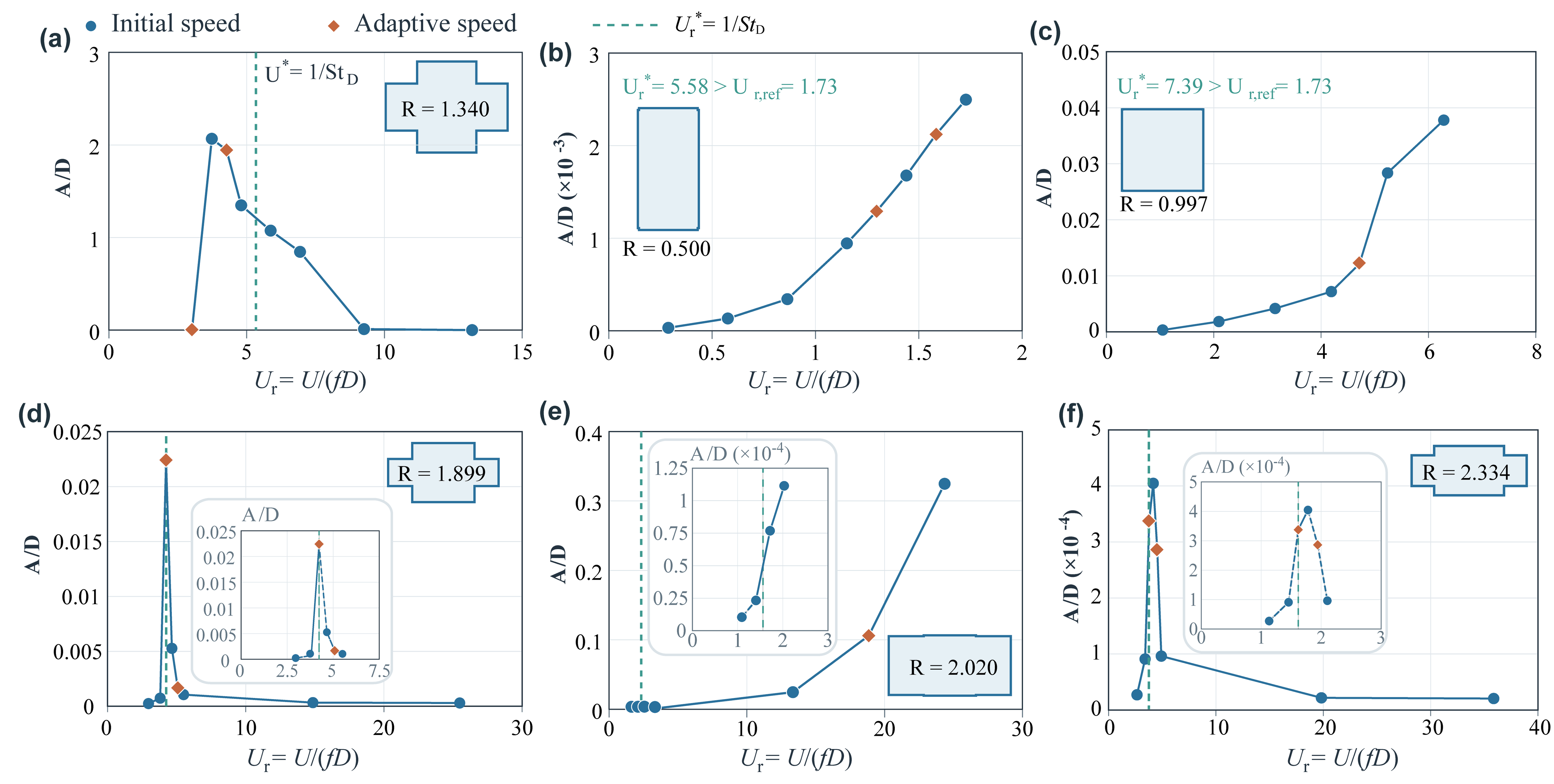}
\caption{Representative sectional vibration responses obtained using adaptive wind-speed sampling: (a) \textit{R}=1.340; (b) \textit{R}=0.500; (c) \textit{R}=0.997; (d) \textit{R}=1.899; (e) \textit{R}=2.020; and (f) \textit{R}=2.334. Blue circles and orange diamonds denote initial and adaptively added evaluations, respectively.}\label{fig:16}
\end{figure}

FSI simulations are used to compare the vibration responses of six representative sections, as shown in Fig.~\ref{fig:16}. The normalized amplitude \textit{A/D} is plotted against the depth-based reduced velocity \(U_{r}=U/\left( fD\right)\). The response curves are constructed using the adaptive wind-speed sampling strategy introduced in Section~\ref{sec:2.2}, with initial and additional evaluations distinguished by different markers.

Figs.~\ref{fig:16}(a,d,f) exhibit characteristic VIV responses, with the associated vibration ranges well resolved by adaptive wind-speed sampling. Their sampled peak amplitudes differ substantially, reaching approximately 2.07, 2.2\ensuremath{\times}10\textsuperscript{-2}, and 4.05\ensuremath{\times}10\textsuperscript{-4}, respectively. The narrow peaks in Figs.~\ref{fig:16}(d,f) highlight the importance of locally refined sampling for capturing the maximum vibration amplitude. Among these cases, the section in Fig.~\ref{fig:16}(f) maintains a particularly weak response throughout the evaluated wind-speed range.

In contrast, the sections in Figs.~\ref{fig:16}(b,c,e) share a common geometric feature: very small corner-recession. Their amplitudes increase throughout the evaluated wind-speed range, without a distinct localized VIV peak, and exhibit a galloping growth trend. This behavior is particularly pronounced in Fig.~\ref{fig:15}(e), where the initially small amplitudes increase substantially at higher wind speeds.

These comparisons emphasize that suppressing galloping does not necessarily reduce the peak VIV response. Comparing Figs.~\ref{fig:16}(a,d,f) with Figs.~\ref{fig:16}(b,c,e) suggests that more pronounced corner-recession can weaken galloping-like response growth while retaining localized VIV peaks. Nevertheless, suitable combinations of sectional proportions and corner-recess dimensions can substantially reduce vibration amplitudes, producing the very weak response observed in Fig.~\ref{fig:16}(f) throughout the investigated wind-speed range. This distinction between galloping suppression and VIV mitigation is consistent with previous studies on corner-cut and rounded square cylinders \citep{Choi1999,Zhao2019}.

\subsection{Structural performance}
\label{sec:4.4}

\subsubsection{Sectional bending stiffness and compression--bending stability}
\label{sec:4.4.1}

\begin{figure}[tbp]
\centering
\includegraphics[width=0.87\linewidth,height=0.73\textheight,keepaspectratio]{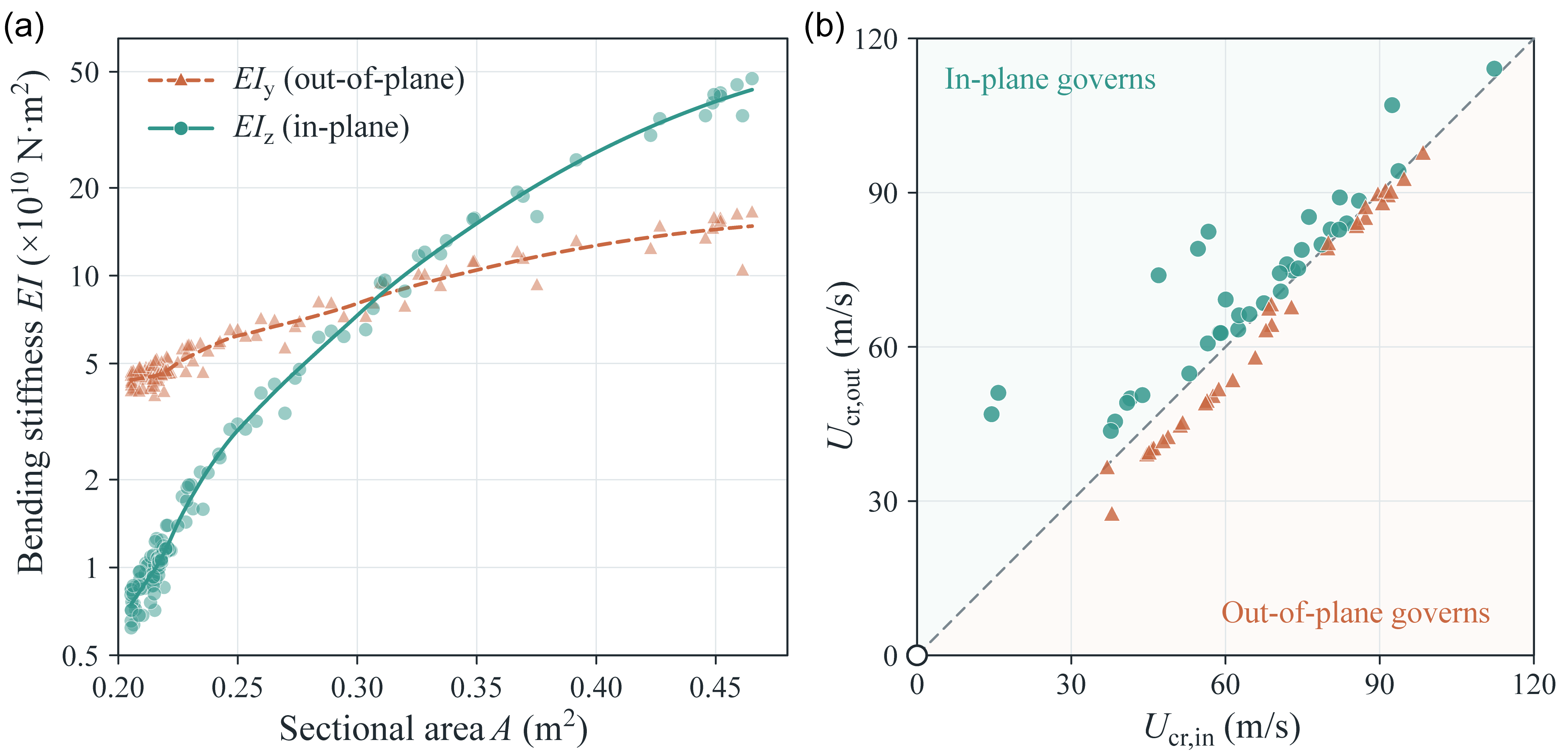}
\caption{Sectional bending stiffness and compression-bending stability: (a) bending stiffnesses versus sectional area; (b) in-plane and out-of-plane critical wind speeds.}\label{fig:17}
\end{figure}

Variations in cross-sectional geometry affect both the sectional bending stiffness and the critical wind speeds for compression-bending stability. Fig.~\ref{fig:17} presents the bending stiffnesses of the sampled sections and compares the critical wind speeds for in-plane and out-of-plane instability.

Fig.~\ref{fig:17}(a) shows that both \textit{EI}\textsubscript{y} and \textit{EI}\textsubscript{z} generally increase with sectional area. At smaller areas, the out-of-plane stiffness \textit{EI}\textsubscript{y} exceeds the in-plane stiffness \textit{EI}\textsubscript{z}. The two fitted curves intersect at approximately \textit{A}=0.30 m\textsuperscript{2}, corresponding to an overall aspect ratio close to \textit{R}=1. Beyond this intersection, \textit{EI}\textsubscript{z} increases more rapidly and becomes greater than \textit{EI}\textsubscript{y}. For the relatively small corner recesses considered here, the overall aspect ratio \textit{R} primarily controls the directional stiffness balance, while the corner recession further modify the local material distribution. The corresponding sectional properties are assigned to the structural finite-element model to calculate internal-force demands, which are subsequently used with the interaction criteria in Section~\ref{sec:3.2} to determine the critical wind speeds. Fig.~\ref{fig:17}(b) compares the in-plane and out-of-plane critical wind speeds at the governing control section of each illustrated design. Points below the diagonal satisfy \(U_{\text{cr},\text{out}}<U_{\text{cr},\text{in}}\) and are governed by the out-of-plane criterion, whereas those above the diagonal are governed by the in-plane criterion. These limits reflect the combined effects of sectional resistance and aerodynamic loading. The distribution of points on both sides of the diagonal shows that either direction can govern, highlighting the need to evaluate both in-plane and out-of-plane compression-bending stability.

\subsubsection{Static deformation}
\label{sec:4.4.2}

\begin{figure}[tbp]
\centering
\includegraphics[width=0.80\linewidth,height=0.60\textheight,keepaspectratio]{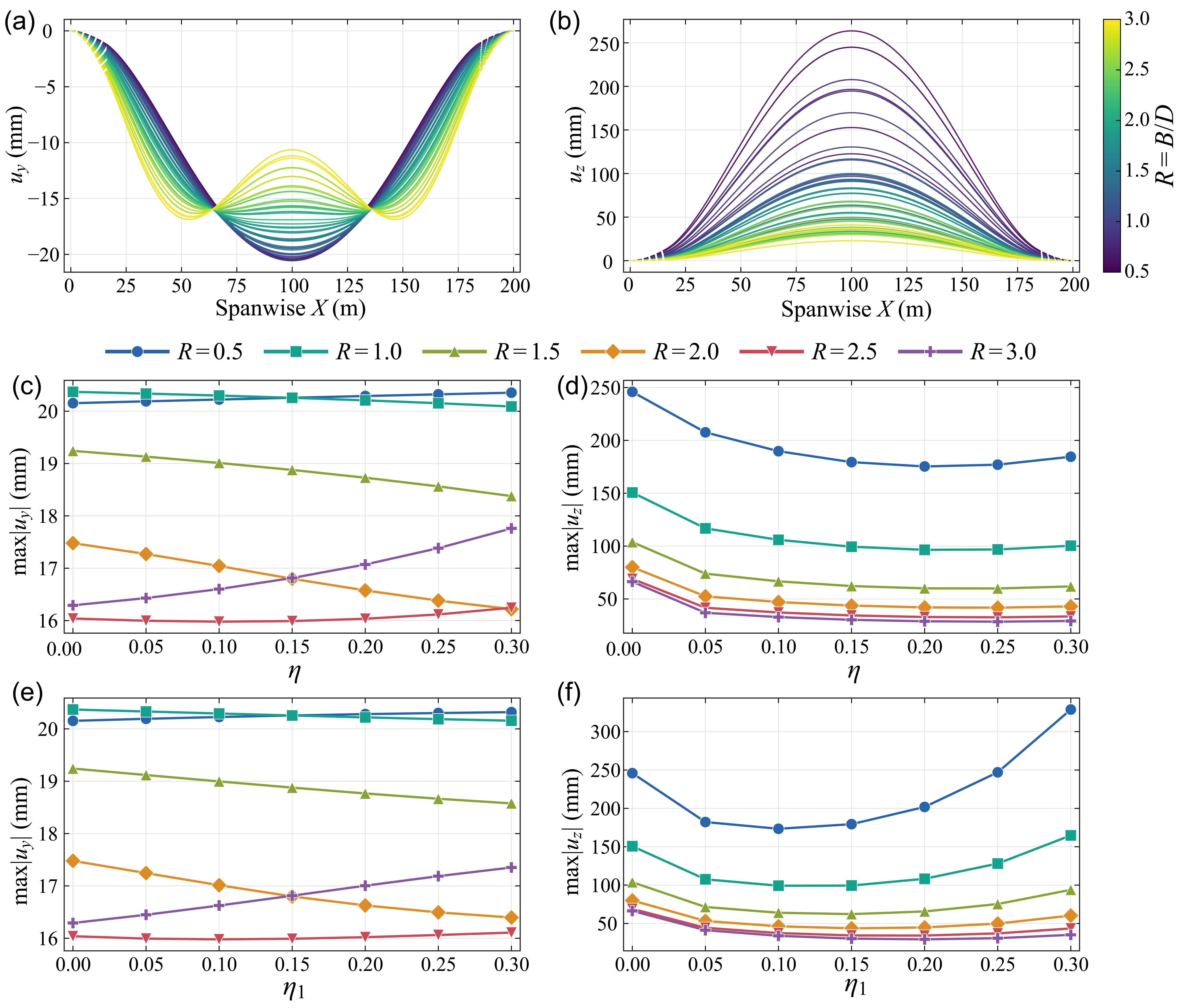}
\caption{Static deformation and effects of corner-recess parameters: (a) vertical displacement; (b) lateral displacement under gravity and transverse wind loading; (c,e) maximum absolute vertical displacement; (d,f) maximum lateral displacement. Panels (c,d) fix \(\eta _{1}=0.15\), while panels (e,f) fix \(\eta =0.15\).}\label{fig:18}
\end{figure}

A static analysis is conducted to examine how sectional geometry affects the deformation of the arch under self-weight and transverse wind loading. Fig.~\ref{fig:18}(a,b) presents the displacement distributions of the 48 Stage-I designs, while Figs.~\ref{fig:18}(c--f) show controlled parameter sweeps at six aspect ratios. The wind-loading comparisons use a common prototype speed of 30 m/s. The vertical displacement \(u_{y}\) represents the gravity response, whereas \(u_{z}\) represents the lateral response to the applied drag loading.

The vertical displacement profiles in Fig.~\ref{fig:18}(a) vary in both magnitude and shape. Lower-\textit{R} designs generally exhibit their largest downward displacement near the crown, whereas higher-\textit{R} designs develop two off-crown displacement maxima. Across the 48 designs, the maximum absolute vertical displacement ranges from 15.98 to 20.56 mm. The lateral profiles in Fig.~\ref{fig:18}(b), by contrast, retain a predominantly single-peaked shape centered near the crown, with maxima ranging from 22.95 to 263.71 mm. Higher-\textit{R} designs generally exhibit smaller lateral displacements under the adopted section-dependent wind loads.

Fig.~\ref{fig:18}(c,e) show that the influence of corner recession on gravity deformation depends on the aspect ratio. Increasing either corner-recess parameter generally reduces the maximum vertical displacement at \textit{R} = 1.5 and \textit{R} = 2.0, but increases it at \textit{R} = 3.0; the \textit{R} = 2.5 curves remain comparatively flat. Fig.~\ref{fig:18}(d,f) show the lateral displacements. Increasing \(\eta\) initially reduces displacement, followed by flattening or a modest increase. The effect of \(\eta _{1}\) is more distinctly non-monotonic, with a pronounced increase beyond intermediate values at low \textit{R}. These trends reflect the combined effects of aerodynamic loading and structural flexibility.

\subsection{Engineering discussion}
\label{sec:4.5}

The six observed Pareto designs offer different balances between vibration suppression and structural stability. As summarized in Table~\ref{tab:4}, all six satisfy the adopted structural stability requirement and the quasi-steady galloping screening condition. Engineering selection should therefore consider the allowable vibration response, required stability and material consumption of the specific project. Sectional area is introduced here as an additional selection criterion to account for material economy.

Design 0105 is a candidate when vibration suppression takes priority, providing the lowest sampled aerodynamic objective, \textit{J}\textsubscript{1}=1.550\ensuremath{\times}10\textsuperscript{-4}, together with a structural indicator of \textit{J}\textsubscript{2} =2.5589. At the other end of the front, design 0015 provides the largest structural indicator, \textit{J}\textsubscript{2} =4.7603, but its vibration objective increases to \textit{J}\textsubscript{1}=0.324, and its sectional area is approximately 5.7\% larger than that of design 0105. Design 0015 therefore illustrates the potential structural benefit of accepting a larger vibration response and greater material consumption. Its suitability depends on whether that response meets the project-specific vibration limit.

Design 0101 provides a practical balance between them with \textit{J}\textsubscript{1} = 4.046\ensuremath{\times}10\textsuperscript{-4} (Fig.~\ref{fig:16}(f)) and \textit{J}\textsubscript{2}=2.9584. Relative to design 0105, it increases the structural indicator by 15.6\% with only 0.9\% more sectional area. It also requires approximately 4.5\% less sectional area than design 0015. These comparisons make design 0101 a relevant candidate when a low vibration response and an increased structural margin are both desired. This designation expresses an engineering preference within the evaluated set rather than a universally optimal choice.

Material economy may lead to a different selection if several designs already satisfy the required vibration limit. For example, design 0095 has a 1.3\% higher structural indicator and a 1.7\% smaller sectional area than design 0101, although its sampled vibration objective is larger. If both responses are acceptable, design 0095 may be preferable on structural and material grounds. Under the same material density, arch geometry and strengthening arrangement, sectional area provides a consistent proxy for steel consumption. However, it does not represent total construction cost, which also depends on fabrication complexity,\allowbreak{} welding and erection requirements. The relatively small area differences among several candidates should therefore be interpreted alongside these practical considerations.

\begin{table}[tbp]
\centering
\caption{Comparison of the six Pareto designs for engineering selection.}\label{tab:4}
\footnotesize\setlength{\tabcolsep}{2pt}
\begin{tabularx}{\linewidth}{@{}lCCCCCC@{}}
\toprule
Parameters & 0105 & 0097 & 0101 & 0095 & 0087 & 0015 \\
\midrule
\textit{R=B/D} & 2.4034 & 2.4455 & 2.3341 & 2.4747 & 2.4511 & 2.0196 \\
\textit{\ensuremath{\eta}} & 0.2231 & 0.2602 & 0.1966 & 0.1593 & 0.2165 & 0.2861 \\
\textit{\ensuremath{\eta}}\textsubscript{1} & 0.1731 & 0.1480 & 0.1898 & 0.1581 & 0.1391 & 0.0137 \\
\textit{A}\textsubscript{sec} (m\textsuperscript{2}) & 0.21820 & 0.21709 & 0.22013 & 0.21633 & 0.21694 & 0.23054 \\
\textit{J}\textsubscript{1} & 1.550\ensuremath{\times}10\textsuperscript{-4} & 2.278\ensuremath{\times}10\textsuperscript{-4} & 4.046\ensuremath{\times}10\textsuperscript{-4} & 5.333\ensuremath{\times}10\textsuperscript{-3} & 4.170\ensuremath{\times}10\textsuperscript{-2} & 3.247\ensuremath{\times}10\textsuperscript{-1} \\
\textit{J}\textsubscript{2} & 2.5589 & 2.5746 & 2.9584 & 2.9968 & 3.5937 & 4.7603 \\
\textit{U}\textsubscript{cr} (m/s) & 79.96 & 80.46 & 92.45 & 93.65 & 112.30 & 148.76 \\
\textit{H} & 9.05 & 5.41 & 6.40 & 10.25 & 11.16 & 3.20 \\
Recommend & Vibration priority & --- & Overall compromise & --- & --- & Stability priority \\
\bottomrule
\end{tabularx}
\end{table}

\section{Conclusions}
\label{sec:5}

This study developed an automated aero-structural optimization framework for the sectional design of steel arches exposed to wind during construction. Parametric geometry generation, sectional CFD--FSI evaluation, structural finite-element analysis, and multi-objective Bayesian optimization were connected within a modular workflow. Its application to a single bare steel arch with a span of 200 m and a rise of 30 m demonstrated how aerodynamic vibration and structural stability can be considered jointly in section design. The main conclusions are as follows.

Adaptive wind-speed sampling improved the resolution of localized response peaks while reducing the cost of the representative CFD--FSI evaluation. Starting from six initial wind speeds, two refinement rounds produced ten evaluated speeds, compared with eighteen in the uniform scan. The adaptive procedure captured a narrow response peak missed by the uniform scan and reduced MPI solver consumption from 230.29 to 69.24 core-hours, a saving of 69.9\%.

The optimization improved the observed aero-structural trade-off. The analysis dataset comprised 123 designs, including 48 initial Latin-hypercube designs and 75 additional designs evaluated over 22 sequential iterations. The reconstructed hypervolume increased from 0.7861 to 0.9843, corresponding to a relative increase of approximately 25.2\%. The final observed Pareto front contained six non-dominated designs, five of which originated from sequential adaptive design sampling.

The steel-arch numerical case demonstrated the effectiveness of the proposed aero-structural optimization framework. Sectional modifications simultaneously altered aerodynamic loading, mass, stiffness, and structural modal properties. Within the examined design space, the parameter combinations of \textit{R=}2.2--2.6, \textit{\ensuremath{\eta}}\textit{=}0.20--0.23, and \textit{\ensuremath{\eta}}\textsubscript{1}\textit{=}0.15--0.17 yield optimal aerodynamic mitigation, while those with \textit{R=}1.9--2.2, \textit{\ensuremath{\eta}=}0.18--0.21, and \textit{\ensuremath{\eta}}\textsubscript{1}\textit{=}0.14--0.16 provide superior structural stability\textit{.} The observed Pareto front clearly confirms that aggressive vibration suppression does not necessarily translate into a higher structural stability margin. As an engineering compromise, design 0101 achieved \textit{J}\textsubscript{1} = 4.046\ensuremath{\times}10\textsuperscript{-4} and \textit{J}\textsubscript{2}=2.9584. Compared with the vibration-priority design 0105, it increased the structural indicator by 15.6\% with only 0.9\% more sectional area.

Based on open-source tools including OpenFOAM, OpenSees, and Dakota, the proposed framework adopts a modular architecture that facilitates further development and customization. Future work could examine selected designs using three-dimensional simulations and wind-tunnel tests, extend the design space to other sectional forms and incorporate multiple wind fields.

\FloatBarrier

\section*{Data availability}

Data will be made available on request.

\appendix
\setcounter{equation}{0}
\renewcommand{\theequation}{\Alph{section}.\arabic{equation}}

\section{Sectional properties and derivation of the structural stability indicator}
\label{sec:A}

\subsection{Sectional properties}
\label{sec:A.1}

The sectional material domain is constructed by subtracting the inner corner-recessed contour from the outer contour, as shown in Fig.8. The outer contour has overall dimensions \textit{B} and \textit{D}, with four rectangular corner recesses of dimensions \textit{a} and \textit{b}. Within the prescribed design domain, the inner contour retains the same corner recession, while its overall width and depth are \textit{B}-2\textit{t}\textsubscript{w} and \textit{D}-2\textit{t}\textsubscript{w}, respectively. The resulting sectional area is

\begin{equation}
A_{\text{sec}}=(BD-4ab)-\left[ (B-2t_{w})(D-2t_{w})-4ab\right] =2t_{w}(B+D-2t_{w})
\label{eq:A.1}
\end{equation}

Let \textit{u} and \textit{v} denote the sectional coordinates along the width and depth, respectively. Applying the parallel-axis theorem to the four removed corner rectangles gives the second moments of the solid region enclosed by the outer contour:

\begin{equation}
\begin{aligned}& I_{\text{in}}^{(\text{o})}=\frac{BD^{3}}{12}-4\left[ \frac{ab^{3}}{12}+ab\left( \frac{D-b}{2}\right) ^{2}\right] , \\ & I_{\text{out}}^{(\text{o})}=\frac{DB^{3}}{12}-4\left[ \frac{ba^{3}}{12}+ab\left( \frac{B-a}{2}\right) ^{2}\right]\end{aligned}
\label{eq:A.2}
\end{equation}

The corresponding inner-contour quantities are obtained by replacing \textit{B} and \textit{D} with \textit{B}-2\textit{t}\textsubscript{w} and \textit{D}-2\textit{t}\textsubscript{w}, while retaining \textit{a} and \textit{b}. Subtracting the inner-contour contributions and simplifying yields

\begin{equation}
\begin{aligned}& I_{\text{in}}=\frac{BD^{3}-(B-2t_{w})(D-2t_{w})^{3}}{12}-4t_{w}ab(D-t_{w}-b), \\ & I_{\text{out}}=\frac{DB^{3}-(D-2t_{w})(B-2t_{w})^{3}}{12}-4t_{w}ab(B-t_{w}-a)\end{aligned}
\label{eq:A.3}
\end{equation}

Equations~\eqref{eq:A.1}--(A.3) show that, under the adopted contour construction, varying the notch dimensions changes the second moments of area without changing \textit{A}\textsubscript{sec} at fixed \textit{B}, \textit{D} and \textit{t}\textsubscript{w}. The elastic section moduli and radii of gyration are subsequently calculated as

\begin{equation}
W_{\text{in}}=\frac{2I_{\text{in}}}{D},\quad W_{\text{out}}=\frac{2I_{\text{out}}}{B},\quad i_{d}=\sqrt{\frac{I_{d}}{A_{\text{sec}}}},\;\;\;d\in \{\text{in},\text{out}\}.
\label{eq:A.4}
\end{equation}

The analytical expressions above describe the geometric relationships underlying the sectional properties. In the implementation, the material domain is constructed using the Python package \textit{Shapely} and discretized for analysis with \textit{sectionproperties}. The sectional area and centroidal second moments of area are obtained through \textit{calculate\_geometric\_}\textit{properties}\textit{(}\textit{)}, whereas the Saint-Venant torsion constant \textit{J} is obtained through the finite-element warping analysis performed by \textit{calculate\_warping\_properties}\textit{()}. The section moduli and radii of gyration are then evaluated using Eq.~\eqref{eq:A.4}. The resulting properties provide the sectional inputs to the OpenSeesPy model and the compression-bending stability assessment.

\subsection{Slenderness parameters and stability reduction factors}
\label{sec:A.2}

The theoretical background for steel-member stability and compressive-bending-member stability is discussed by \citet{Shen2005} and Li (1992). The present assessment adopts effective-length approximations to determine the slenderness parameters and reference elastic stability forces for the in-plane and out-of-plane directions. These quantities serve as inputs to the beam-column interaction criteria

For the parabolic arch with span \textit{L} and rise \textit{h}, the half-arch length is

\begin{equation}
S_{1/2}=\int_{0}^{L/2} \sqrt{1+\left[ \frac{4h(L-2X)}{L^{2}}\right] ^{2}}\,\text{d}X
\label{eq:A.5}
\end{equation}
where \textit{X} is the horizontal coordinate measured from the springing. In the implementation, this integral is evaluated using a polygonal approximation with 4000 segments over the half-span. The in-plane slenderness ratio and reference elastic stability force are defined as

\begin{equation}
\lambda _{\text{in}}=\frac{\mu _{\text{in}}S_{1/2}}{i_{\text{in}}},\quad N^{\prime}_{\text{E},\text{in}}=\frac{\pi ^{2}EA_{\text{sec}}}{1.1\lambda _{\text{in}}^{2}},\quad \mu _{\text{in}}=0.67
\label{eq:A.6}
\end{equation}

For out-of-plane assessment, the adopted arch approximation gives

\begin{equation}
N^{\prime}_{\text{E},\text{out}}=\frac{\pi ^{2}EI_{\text{out}}}{\mu _{\text{out}}^{2}L^{2}\left[ 1+\frac{8}{3}\left( \frac{h}{L}\right) ^{2}\right] ^{2}},\quad \lambda _{\text{out}}=\pi \sqrt{\frac{EA_{\text{sec}}}{N^{\prime}_{\text{E},\text{out}}}},\quad \mu _{\text{out}}=0.70.
\label{eq:A.7}
\end{equation}

For either direction \(d\in \{\text{in},\text{out}\}\), the normalized slenderness is

\begin{equation}
\bar{\lambda}_{d}=\frac{\lambda _{d}}{\pi}\sqrt{\frac{f_{y}}{E}},
\label{eq:A.8}
\end{equation}
where \(f_{y}\) is the prescribed yield stress. The axial-compression reduction relation used in the evaluator is

\begin{equation}
\hat{\varphi}_{d}=\left\{ \begin{array}{ll}1-0.65\bar{\lambda}_{d}^{2}, & \bar{\lambda}_{d}\leq 0.215, \\ \frac{Q_{d}-\sqrt{Q_{d}^{2}-4\bar{\lambda}_{d}^{2}}}{2\bar{\lambda}_{d}^{2}}, & \bar{\lambda}_{d}>0.215,\end{array}\right. \quad Q_{d}=0.965+0.300\bar{\lambda}_{d}+\bar{\lambda}_{d}^{2}.
\label{eq:A.9}
\end{equation}

This correspondence concerns the reduction-curve coefficients; the applicability of the selected curve and the effective-length assumptions to the corner-recessed arch section remains part of the modeling assumptions. For numerical evaluation, the reduction factors are bounded as

\begin{equation}
\varphi _{d}=min\,\left[ 1,\,max\,\left( 0.05,\hat{\varphi}_{d}\right)\right] .
\label{eq:A.10}
\end{equation}

\subsection{Equivalent wind-induced bending moments}
\label{sec:A.3}

For each control location \textit{s}, the gravity-induced compressive-force magnitude \textit{N}\textsubscript{\textit{s}} and in-plane bending-moment magnitude \textit{M}\textsubscript{G,}\textsubscript{\textit{s}} are obtained from the element-end force envelopes described in Section~\ref{sec:3.2.3}. Wind effects are introduced separately through equivalent bending moments.

The equivalent drag and lift forces per unit length are

\begin{equation}
q_{\text{D}}(U)=\frac{1}{2}\rho _{\text{a}}DC_{\text{D}}U^{2},\quad q_{\text{L}}(U)=\frac{1}{2}\rho _{\text{a}}BC_{\text{L}}U^{2}
\label{eq:A.11}
\end{equation}
where \(\rho _{\text{a}}\) is the air density and positive lift acts upward. Introducing an equivalent moment factor \(\xi\), the wind-moment coefficients are

\begin{equation}
\alpha _{\text{out}}=\frac{1}{2}\xi \rho _{\text{a}}DC_{\text{D}}L^{2},\quad \alpha _{\text{in}}=\frac{1}{2}\xi \rho _{\text{a}}BC_{\text{L},B}L^{2}.
\label{eq:A.12}
\end{equation}

The adopted value \(\xi =1/12\) is motivated by the fixed-end moment magnitude \(qL^{2}/12\) for a straight beam under a uniformly distributed transverse load. Here, it is used as a simplified load-to-moment conversion factor, not as an exact solution for the arch moment distribution.

The wind-speed-dependent internal-force quantities entering the interaction conditions are therefore

\begin{equation}
M_{\text{out},s}(U)=\alpha _{\text{out}}U^{2},M_{\text{in},s}(U)=M_{\text{G},s}-\alpha _{\text{in}}U^{2},N_{s}(U)=N_{s}.
\label{eq:A.13}
\end{equation}

The aerodynamic coefficients remain fixed during the wind-speed inversion, and the same wind-moment coefficients are applied at the three control locations. Thus, Eq.~\eqref{eq:A.13} represents a reduced-order loading approximation rather than a finite-element reanalysis at every wind speed.

\subsection{Inversion of the interaction criteria and evaluation of critical wind speeds}
\label{sec:A.4}

For each control location \textit{s} and stability direction \textit{d}, the moment-amplification denominator is

\begin{equation}
\Delta _{s,d}=1-\frac{0.8N_{s}}{N^{\prime}_{\text{E},d}},\quad d\in \{\text{in},\text{out}\}.
\label{eq:A.14}
\end{equation}

Substituting Eq.~\eqref{eq:A.13} into the two compression-bending interaction conditions in Section~\ref{sec:3.2.3} gives

\begin{equation}
r_{s,d}(U)=r_{s,d}(0)+\frac{\mathcal{K}_{s,d}}{f_{d}}U^{2}\leq 1,\quad \mathcal{C}_{s,d}=f_{d}\left[ 1-r_{s,d}(0)\right] ,
\label{eq:A.15}
\end{equation}
where \(k\in \{\text{in},\text{out}\}\), \(f_{d}\) is the prescribed stress limit, \(\mathcal{K}_{s,d}\) is the coefficient of the wind-speed-squared term, and \(\mathcal{C}_{s,d}\) is the initial stress margin. The interaction condition can therefore be written as \(\mathcal{K}_{s,d}U^{2}\leq \mathcal{C}_{s,d}\).

For the in-plane interaction condition, the coefficients are

\begin{equation}
\begin{aligned}& \mathcal{K}_{s,\text{in}}=\frac{\chi \beta _{t,\text{out}}\alpha _{\text{out}}}{\varphi _{b}W_{\text{out}}}-\frac{\beta _{m,\text{in}}\alpha _{\text{in}}}{\gamma _{\text{in}}W_{\text{in}}\Delta _{s,\text{in}}}, \\ & \mathcal{C}_{s,\text{in}}=f_{d}-\frac{N_{s}}{\varphi _{\text{in}}A_{\text{sec}}}-\frac{\beta _{m,\text{in}}M_{\text{G},s}}{\gamma _{\text{in}}W_{\text{in}}\Delta _{s,\text{in}}}.\end{aligned}
\label{eq:A.16}
\end{equation}

For the out-of-plane interaction condition,

\begin{equation}
\begin{aligned}& \mathcal{K}_{s,\text{out}}=\frac{\beta _{m,\text{out}}\alpha _{\text{out}}}{\gamma _{\text{out}}W_{\text{out}}\Delta _{s,\text{out}}}-\frac{\chi \beta _{t,\text{in}}\alpha _{\text{in}}}{\varphi _{b}W_{\text{in}}}, \\ & \mathcal{C}_{s,\text{out}}=f_{d}-\frac{N_{s}}{\varphi _{\text{out}}A_{\text{sec}}}-\frac{\chi \beta _{t,\text{in}}M_{\text{G},s}}{\varphi _{b}W_{\text{in}}}.\end{aligned}
\label{eq:A.17}
\end{equation}

For finite inputs and positive stability reduction factors, the corresponding directional critical speed is evaluated as

\begin{equation}
U_{s,d}=\left\{ \begin{array}{ll}0, & \Delta _{s,d}\leq 0\;\text{or }\mathcal{C}_{s,d}\leq 0, \\ \sqrt{\frac{\mathcal{C}_{s,d}}{\mathcal{K}_{s,d}}}, & \Delta _{s,d}>0,\;\;\;\mathcal{C}_{s,d}>0,\;\;\;\mathcal{K}_{s,d}>0, \\ +\infty , & \Delta _{s,d}>0,\;\;\;\mathcal{C}_{s,d}>0,\;\;\;\mathcal{K}_{s,d}\leq 0.\end{array}\right.
\label{eq:A.18}
\end{equation}

A zero value indicates that the adopted criterion has no positive initial margin or has a non-positive amplification denominator. An infinite value indicates that the fixed-coefficient interaction expression has no finite positive root; it does not imply unlimited physical wind resistance.

The governing critical wind speed is the minimum over the prescribed control locations and both interaction directions. The structural objective is then obtained by normalizing this speed by the reference wind speed:

\begin{equation}
U_{\text{cr}}(x)=\operatorname*{min}\limits_{s\in \mathcal{S}}\operatorname*{min}\limits_{d\in \{\text{in},\text{out}\}}U_{s,d}(x),\quad \mathcal{S}=\left\{ 0,\frac{L}{4},\frac{L}{2}\right\}
\label{eq:A.19}
\end{equation}

\begin{equation}
J_{2}(x)=\frac{U_{\text{cr}}(x)}{U_{\text{ref}}}
\label{eq:A.20}
\end{equation}

The control locations correspond to the springing, quarter-span, and crown of the modeled half-arch.

\FloatBarrier
\setcounter{figure}{0}
\renewcommand{\thefigure}{B.\arabic{figure}}
\renewcommand{\theHfigure}{B.\arabic{figure}}

\section{Leave-one-out cross-validation of reconstructed GP surrogates}
\label{sec:B}

\begin{figure}[tbp]
\centering
\includegraphics[width=0.67\linewidth,height=0.73\textheight,keepaspectratio]{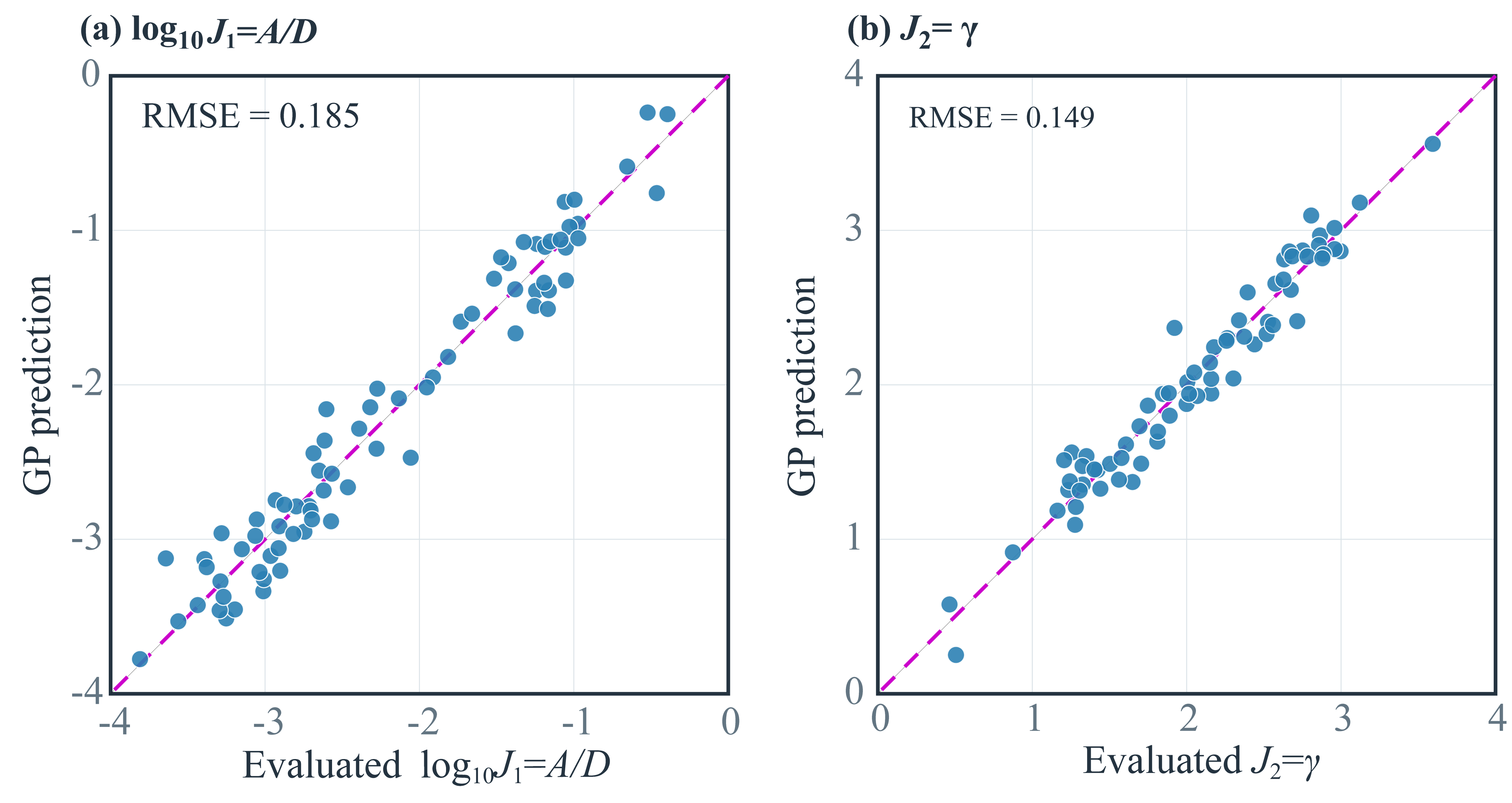}
\caption{Evaluated responses versus leave-one-out predictions for Stage-II designs: (a) aerodynamic objective; (b) structural objective.}\label{fig:B.1}
\end{figure}

The predictive accuracy of the reconstructed Gaussian-process (GP) surrogates was assessed using leave-one-out cross-validation (LOOCV). Each observation was withheld in turn, and response standardization and hyperparameter fitting were repeated using only the remaining training data. Fig.~\ref{fig:B.1} presents the held-out predictions for the Stage-II designs, comprising 75 observations. Predictions are obtained by refitting the reconstructed GP models to the remaining valid observations in the full analysis dataset. Dashed lines indicate ideal agreement. The aerodynamic objective was evaluated in \(log_{10}(J_{1})\) space, whereas the structural objective was evaluated in \(J_{2}\) space. The results indicate that the GP model has some predictive capability.

\end{document}